\documentclass[aps,showpacs,pra,superscriptaddress,preprintnumbers,amsmath,amssymb,twocolumn,longbibliography]{revtex4-2}
\usepackage[utf8]{inputenc}
\usepackage[T1]{fontenc}
\usepackage{graphicx}
\usepackage{dcolumn}
\usepackage{appendix}
\usepackage{amsmath}
\usepackage{bm}
\usepackage[dvipsnames]{xcolor}
\usepackage{multirow}
\usepackage{pifont}
\newcommand{\relaxket}[1]{\lvert{#1}\rangle}

\usepackage{subfigure}
\usepackage{bbm}
\usepackage{enumerate}

\usepackage[
bookmarks=true,
colorlinks,
linkcolor=blue,
urlcolor=blue,
citecolor=black,
plainpages=false,
pdfpagelabels,
final,
breaklinks=true
]{hyperref}

\usepackage{titlesec}

\usepackage{array}

\usepackage{physics}

\begin{document}
\allowdisplaybreaks

\title{Emergence of Gaussian entanglement and non-Gaussianity in high-harmonic generation driven by bright squeezed light}

\author{J.~Rivera-Dean}
\email{physics.jriveradean@proton.me}
\email{javier.dean@ucl.ac.uk}
\affiliation{Department of Physics and Astronomy, University College London, Gower Street, London WC1E 6BT, UK}

\author{M.~Even-Tzur}
\affiliation{Max Planck Institute for the Structure and Dynamics of Matter, Hamburg, Germany}

\author{M.~F.~Ciappina}
\affiliation{Department of Physics, Guangdong Technion - Israel Institute of Technology, 241 Daxue Road, Shantou, Guangdong, China, 515063}
\affiliation{Technion - Israel Institute of Technology, Haifa, 32000, Israel}
\affiliation{Guangdong Provincial Key Laboratory of Materials and Technologies for Energy Conversion, Guangdong Technion - Israel Institute of Technology, 241 Daxue Road, Shantou, Guangdong, China, 515063}

\author{C.~Granados}
\affiliation{Eastern Institute of Technology, Ningbo 315200, China}

\author{O.~Cohen}
\affiliation{Technion - Israel Institute of Technology, Haifa, 32000, Israel}
\affiliation{Guangdong Provincial Key Laboratory of Materials and Technologies for Energy Conversion, Guangdong Technion - Israel Institute of Technology, 241 Daxue Road, Shantou, Guangdong, China, 515063}

\author{P.~Stammer}
\affiliation{ICFO -- Institut de Ciencies Fotoniques, The Barcelona Institute of Science and Technology, 08860 Castelldefels (Barcelona)}
\affiliation{Atominstitut, Technische Universität Wien, 1020 Vienna, Austria}

\begin{abstract}
	High harmonic generation (HHG) is a highly nonlinear optical process in which radiation from a strong driving field is up-converted into its high-order harmonics.~In atomic systems, this nonlinearity manifests itself through the intensity scaling of the emitted harmonics with the driving field strength.~Despite the highly nonlinear nature of HHG, when the driving field is prepared in a classical Gaussian state and atomic depletion remains negligible, the quantum statistical properties of the generated harmonics retains classical Gaussian quantum statistics.~Driving HHG with bright squeezed vacuum (BSV) light challenges this paradigm, as its enhanced field fluctuations can modify the statistical properties of the generated harmonics.~In this work, we investigate the conditions under which BSV-driven HHG gives rise to non-classical Gaussian states, and identify the regimes where this Gaussian description breaks down.~For bichromatic driving by a strong coherent field at frequency $\omega$ and a perturbative BSV field at $2\omega$, the even-harmonic response is approximately linear in the BSV quadrature, leading to non-classical multimode Gaussian entanglement in the harmonic field.~We show that this state can be described as a distributed collective squeezed mode over the even-harmonic manifold, and characterize its covariance matrix, entanglement structure, and quantum teleportation fidelity as an operational benchmark.~Our results highlight the potential of non-classically driven HHG as a platform for engineering Gaussian and non-Gaussian states of light in the extreme ultraviolet regime.
\end{abstract}

\maketitle

\section{INTRODUCTION}
Gaussian distributions are among the most widely used statistical distributions, owing both to their mathematical tractability and to their natural emergence across a broad range of systems.~Their convenience stems from the fact that they are entirely characterized by their mean and variance, with all higher-order moments vanishing identically, whereas their prevalence across a vast range of systems of different nature and composition arises naturally, often as a consequence of the central limit theorem~\cite{jaynes_central_2003}.~This ubiquity extends to quantum optics, where Gaussian structures emerge in a variety of contexts, ranging from operational settings~\cite{serafini_gaussian_2023} to the statistical properties of optical fields themselves~\cite{ScullyBookCh2,GerryBookCh7}.~Beyond their mathematical convenience, such Gaussian structures can be exploited in different ways, for instance directly in quantum information protocols~\cite{braunstein_quantum_2005,usenko_continuous-variable_2026}, or as building blocks for the generation of more complex non-Gaussian states~\cite{ourjoumtsev_generating_2006,neergaard-nielsen_generation_2006,ourjoumtsev_generation_2007}.

The operational convenience of Gaussian states does not imply, however, classicality or triviality; on the contrary, the generation and control of non-classical Gaussian states of light remains a central objective in quantum optics~\cite{braunstein_quantum_2005,usenko_continuous-variable_2026}.~From a physical perspective, such states are typically produced via nonlinear optical interactions.~For example, squeezed states of light~\cite{walls_squeezed_1983}, known for their capability of yielding below vacuum-noise fluctuations in certain measurements~\cite{xiao_precision_1987,abadie_gravitational_2011}, can be generated through processes such as four-wave mixing~\cite{slusher_observation_1985}, and optical parametric oscillation~\cite{wu_squeezed_1987,lam_optimization_1999} or amplification~\cite{schneider_generation_1998}, where the pump photons are converted into correlated photon pairs under energy conservation.~In the degenerate case, these pairs share the same frequency and spatial mode, yielding single-mode squeezed light; when degeneracy is not enforced, the resulting state corresponds to a two-mode squeezed state~\cite{heidmann_observation_1987}, where the photons are entangled in different spatial or frequency modes~\cite{ou_realization_1992}.~By combining these processes with broadband pump fields~\cite{roslund_wavelength-multiplexed_2014,medeiros_de_araujo_full_2014,gerke_full_2015}, concatenation with linear optical elements~\cite{van_loock_building_2007,chen_experimental_2014} or appropriate shaping of ultrashort pulses acting as local oscillators~\cite{cai_multimode_2017}, it is possible to engineer multipartite Gaussian entangled states.

In contrast, high-harmonic generation (HHG) arises from strong-field light-matter interaction, in which an intense driving field induces ultrafast electron dynamics leading to the emission of high-order harmonics~\cite{krause_high-order_1992,corkum_plasma_1993,lewenstein_theory_1994}.~These harmonics span from the infrared to the extreme ultraviolet~\cite{antoine_attosecond_1996,drescher_x-ray_2001,paul_observation_2001}, and are typically emitted in the form of Gaussian uncorrelated states~\cite{lewenstein_generation_2021,stammer_quantum_2023}, though temporal~\cite{stammer_squeezing_2023,rivera-dean_squeezed_2024} or material~\cite{lange_electron-correlation-induced_2024,theidel_evidence_2024,theidel_observation_2025,lange_hierarchy_2025} correlations can nevertheless yield a multipartite Gaussian entangled state between the harmonic modes.~Alternatively, the emerging possibility of driving HHG with bright squeezed light~\cite{gorlach_high-harmonic_2023,rasputnyi_high_2024} opens new avenues for transferring the non-classical Gaussian properties of the driving field to the generated harmonics~\cite{tzur_generation_2024}.~Under these conditions, the emitted harmonics have been shown to exhibit super-Poissonian statistics~\cite{lemieux_photon_2025,tzur_attosecond-resolved_2025,stammer_weak_2025,petrovic_generation_2026,lyu_attosecond_2026,stammer_fluctuation-induced_2026}, entanglement features~\cite{rivera-dean_attosecond_2026}, and phase-space signatures reminiscent of non-Gaussian states~\cite{tzur_attosecond-resolved_2025}.~These observations point to a rich interplay between the Gaussian statistics of the driving field and the nonlinear strong-field response of the medium.~While the origin of non-Gaussian features in BSV-driven HHG can be understood from the nonlinear dependence of the harmonic emission on the driving-field fluctuations~\cite{tzur_attosecond-resolved_2025}, a systematic characterization of the Gaussian and non-Gaussian regimes, as well as the type of entanglement structure in the outgoing radiation, remains desirable.~Moreover, identifying this boundary is essential for distinguishing HHG as a source of multimode Gaussian entanglement from HHG as a mechanism for generating genuinely non-Gaussian harmonic states.

This work precisely aims at addressing these open questions by identifying conditions under which bright squeezed light driving HHG in atomic media can give rise to non-classical Gaussian states in the generated harmonic field.~We identify bichromatic driving configurations, composed of a strong coherent field at frequency $\omega$ and a weaker, yet bright, squeezed field at $2\omega$, as a scenario in which entangled Gaussian states can emerge.~Employing Gaussian quantum information methods based on the full covariance matrix of the obtained Gaussian states, we analytically characterize the resulting states and quantify multipartite entanglement correlations among the generated harmonics for different atomic systems and collective optical-mode configurations.~We then benchmark whether the established correlations are enough for obtaining a quantum advantage in continuous variable quantum teleportation~\cite{braunstein_teleportation_1998}, a paradigmatic protocol in the context of Gaussian quantum information.~Finally, we determine the conditions under which this configuration induces a transition from Gaussian to non-Gaussian states in the harmonic emission.

Additionally to contributions from previous studies, which have primarily focused on individual harmonic modes and their photon statistics or phase-space properties~\cite{lemieux_photon_2025,tzur_attosecond-resolved_2025,stammer_weak_2025}, and on demonstrating entanglement between the fundamental and harmonics~\cite{rivera-dean_attosecond_2026}, the present work identifies the conditions under which bichromatic BSV driving configurations give rise to non-classical Gaussian states of the emitted harmonic radiation, and addresses its multimode non-classical structure. In particular, we identify the relevant collective modes of the harmonic field, enabling the reconstruction of its full covariance matrix and the characterization of multipartite entanglement across the harmonic spectrum, as well as an assessment of the operational performance of these correlations in continuous-variable quantum teleportation.~We further identify the conditions governing the transition from Gaussian to non-Gaussian harmonic emission.

\section{THEORY BACKGROUND}\label{Sec:ThBck}
We begin our discussion by first establishing the conditions whereby HHG driven by an arbitrary pure quantum state of light in atomic systems and under low-depletion conditions, yields a Gaussian state.~We then focus to bright squeezed drivers, identifying bichromatic configurations as a viable scheme that satisfies these requirements.

\subsection{Conditions for Gaussianity}\label{Sec:Conds:Gauss}
Given a quantum state, its characteristic function encodes all statistical properties in phase-space and therefore completely describes it.~In the multimode case, for a density operator $\hat{\rho}$ acting on a $d$-mode Hilbert space $\mathcal{H}^{\otimes d}$, the characteristic function $\zeta(\boldsymbol{z})$ is defined as~\cite{lai_characteristic_1989,vogel_bosonic_2006,serafini_phase_2023}
\begin{equation}\label{Eq:charc:func}
	\zeta(\boldsymbol{z})
		= \tr[\hat{\rho} \hat{\boldsymbol{D}}(\boldsymbol{z})],
\end{equation}
where $\hat{\boldsymbol{D}}(\boldsymbol{z}) = \bigotimes_{i=1}^d \hat{D}_i(z_i)$ with $\hat{D}_i(z_i) = \mathrm{exp}[z_i\hat{a}_i^\dagger - z_i^*\hat{a}_i]$ the single-mode displacement operator acting on the Hilbert space $\mathcal{H}_i$, and $\hat{a}_i$ ($\hat{a}^\dagger_i$) the corresponding annihilation (creation) operators.~Importantly, $\hat{\rho}$ is said to be Gaussian if and only if its characteristic function has a Gaussian form, in which case it is entirely characterized by its first and second moments.~Consequently, $\zeta(\boldsymbol{z})$ provides a direct means of identifying Gaussian states and of determining the conditions under which a parametrized state exhibits Gaussian behavior.

In this work, we are interested in the properties of the quantum optical state generated after driving HHG in an atomic medium initially prepared in its ground state $\ket{\text{g}}$, using an input quantum optical state $\ket{\Phi(t_0)}$.~For generality, we allow $\ket{\Phi(t_0)}$ to span one or multiple frequency modes, and set all other high-harmonic modes to be in the vacuum state (hereupon referred to as ``HH'').~Consequently, the initial state of the joint light-matter system reads
\begin{equation}\label{Eq:init:state:coh:expansion}
	\ket{\Psi(t_0)}
		= \int \dd^2 \boldsymbol{\alpha}
				\ c(\boldsymbol{\alpha}) \ket{\text{g}}
						\otimes \ket{\boldsymbol{\alpha}}
							\bigotimes_{q \in \text{HH}}\ket{0_q},
\end{equation}
Here, for convenience, we have expanded the initial driving field in the coherent state basis by inserting the identity operator $\mathbbm{1} = \pi^{-d}\bigotimes^d_{q=1} \int \dd^2\alpha_q \dyad{\alpha_q}$, where $d$ denotes the number of frequency modes in the driving field ($\ket{\boldsymbol{\alpha}} = \bigotimes^d_{q=1} \ket{\alpha_q}$), and the probability amplitudes are $c(\boldsymbol{\alpha}) = \pi^{-d}\prod^d_{q=1} \braket{\alpha_q}{\Phi(t_0)}$.~Then, given the time-evolution propagator $\hat{U}(t,t_0)$, the state at time $t\geq t_0$ can be written as
\begin{equation}\label{Eq:joint:state:t}
	\begin{aligned}
	\ket{\Psi(t)}
		&= \hat{U}(t,t_0)\ket{\Psi(t_0)}
		\\&
		= \int \dd^2 \boldsymbol{\alpha}
				\ c(\boldsymbol{\alpha})
					\hat{\boldsymbol{D}}_\text{dr}(\boldsymbol{\alpha})
						\hat{U}(t,t_0;\boldsymbol{\alpha})
							\ket{\text{g}}\otimes \ket{\boldsymbol{0}},
	\end{aligned}
\end{equation}
where we have defined the displaced propagator $\hat{U}(t,t_0;\boldsymbol{\alpha}) = \hat{\boldsymbol{D}}^\dagger_\text{dr}(\boldsymbol{\alpha})\hat{U}(t,t_0)\hat{\boldsymbol{D}}_\text{dr}(\boldsymbol{\alpha})$, with $\hat{\boldsymbol{D}}_\text{dr}(\boldsymbol{\alpha})$ the multimode displacement operator acting on the driving modes (denoted via the subscript ``dr''), and $\ket{\boldsymbol{0}}$ the vacuum state in all modes.

In the context of strong-field interactions, and using atomic units throughout~\footnote{In atomic units, we set $\hbar = m_{\mathsf{e}} = \abs{\mathsf{e}} = k_c = 1$.}, the Hamiltonian governing the light-matter interaction is given by $\hat{H}(t) = \hat{H}_{\text{at}} + \hat{r} \hat{E}(t)$, where $\hat{H}_{\text{at}}$ is the atomic Hamiltonian and $\hat{r} \hat{E}(t)$ describes the interaction in the length gauge and under the dipole approximation.~The electric field operator is given by $\hat{E}(t) = -i \sum_{q} \kappa \sqrt{q}[\hat{a}_{q}e^{-i\omega_q t} + \text{h.c.}]$, where linear polarization is assumed in what follows, and $\kappa$ denotes the light-matter coupling.~Thus, the displaced propagator $\hat{U}(t,t_0;\boldsymbol{\alpha})$ satisfies
\begin{equation}
	i \pdv{\hat{U}(t;\boldsymbol{\alpha})}{t}
		= \big[
				\hat{H}_{\text{at}}
				+ \hat{r}
					\big(
						\hat{E}(t)
						+ E_{\text{cl}}(t;\boldsymbol{\alpha})
					\big) 
			\big]\hat{U}(t;\boldsymbol{\alpha}),
\end{equation}
where $E_{\text{cl}}(t;\boldsymbol{\alpha}) \equiv \text{tr}[\hat{E}(t) \dyad{\boldsymbol{\alpha}}\bigotimes_{q\in \text{HH}}\dyad{0_q}]$ denotes the classical field associated with the coherent state $\ket{\boldsymbol{\alpha}}$. Among all possible strong-field processes~\cite{amini_symphony_2019}, we focus on the parametric process of HHG where the electron returns to the ground state after the interaction.~Projecting Eq.~\eqref{Eq:joint:state:t} onto $\ket{\text{g}}$, the resulting quantum optical state reads
\begin{equation}
	\begin{aligned}
	\ket{\Phi(t)}
		&= \braket{\text{g}}{\Psi(t)}
		\\&
		= \int \dd^2 \boldsymbol{\alpha} \ c(\boldsymbol{\alpha})
				\hat{D}_{\text{dr}}(\boldsymbol{\alpha})
					\mel{\text{g}}{\hat{U}(t,t_0;\boldsymbol{\alpha})}{\text{g}}
						\otimes \ket{\boldsymbol{0}}.
	\end{aligned}
\end{equation}
Under low-depletion conditions~\cite{stammer_squeezing_2023}, the matrix element of the propagator can be approximated as $\mel{\text{g}}{\hat{U}(t,t_0;\boldsymbol{\alpha})}{\text{g}} \simeq \hat{\boldsymbol{D}}(\boldsymbol{\chi}(t;\boldsymbol{\alpha}))$~\cite{lewenstein_generation_2021,stammer_quantum_2023} where $\chi_q(t;\boldsymbol{\alpha}) \equiv \chi_q(\boldsymbol{\alpha}) =-\kappa \sqrt{q} \int_{t_0}^t \dd \tau \langle \hat{r}(\tau;\boldsymbol{\alpha})\rangle e^{i\omega_q t}$, and $ \langle \hat{r}(\tau;\boldsymbol{\alpha})\rangle$ denotes the time-dependent dipole moment~\footnote{In evaluating $\chi_q(\alpha)$, we have used the strong-field approximation, which in general does not accurately reproduce the exact harmonic intensities for low harmonic orders. However, comparison with time-dependent Schrödinger equation results shows that it correctly captures the linear scaling with respect to $E_{2\omega}$, which is the relevant behavior in the present context.} obtained when using $E_{\text{cl}}(t;\boldsymbol{\alpha})$ as driving field~\cite{lewenstein_theory_1994,amini_symphony_2019,RBSFA}.~The quantum optical state after HHG can then be written as
\begin{equation}\label{Eq:QO:HHG:gen}
	\ket{\Phi(t)}
		= \! \int \dd^2 \boldsymbol{\alpha} \
				c(\boldsymbol{\alpha})
					e^{i\varphi_{\text{dr}}(\boldsymbol{\alpha})}
					\ket{ \boldsymbol{\alpha} + \boldsymbol{\chi}_{\text{dr}}(\boldsymbol{\alpha})}
						\!\!\bigotimes_{q\in \text{HH}}\! \ket{\chi_q(\boldsymbol{\alpha})},
\end{equation}
where $\varphi_{\text{dr}}(\boldsymbol{\alpha}) = \sum^d_{q=1} \text{Im}[ \chi_{q}(\boldsymbol{\alpha}) \alpha_q^*]$.

With the above results, combining Eqs.~\eqref{Eq:charc:func} and \eqref{Eq:QO:HHG:gen}, the characteristic function of the quantum optical state generated after HHG reads
\begin{equation}
	\begin{aligned}
	\zeta(\boldsymbol{z})
		&= \int \dd^2 \boldsymbol{\alpha}
			\! \int \dd^2 \boldsymbol{\beta}\
				c(\boldsymbol{\alpha})
					c^*(\boldsymbol{\beta})
						e^{i[\varphi_{\text{dr}}(\boldsymbol{\alpha}) 
								- \varphi_{\text{dr}}^*(\boldsymbol{\beta})]}
		\\& \hspace{1cm}\times
				e^{\sum_q \varphi_q(z_q,\boldsymbol{\alpha})}
				\prod_{q\in \text{HH}}
					\braket{\chi_q(\boldsymbol{\beta})}{\chi_q(\boldsymbol{\alpha}) + z_q}
		\\&\hspace{1cm}\times
		\braket{\boldsymbol{\beta} + \boldsymbol{\chi}_{\text{dr}}(\boldsymbol{\beta})}{\boldsymbol{\alpha} + \boldsymbol{\chi}_{\text{dr}}(\boldsymbol{\alpha}) + \boldsymbol{z}_{\text{dr}}},
	\end{aligned}
\end{equation}
where $\varphi_q(z_q,\boldsymbol{\alpha}) = \text{Im}[z_q (\alpha^*_q \delta_{q\in\text{dr}}+\chi^*_q(\boldsymbol{\alpha}))]$.~Since both coherent state overlaps and the phase factors $\varphi(\cdot)$ are Gaussian functions of their arguments, the Gaussian character of $\zeta(\boldsymbol{z})$ is determined by the dependence of the exponent on the integration variables $\boldsymbol{\alpha}$ and $\boldsymbol{\beta}$, as well as the functional form of $c(\boldsymbol{\alpha})$.~Thus, a simple sufficient route for the resulting quantum optical state to remain Gaussian is provided by the following two conditions:
\begin{enumerate}[(1)]
	\item $c(\boldsymbol{\alpha})$ must be Gaussian;
	\item $\boldsymbol{\chi}(\boldsymbol{\alpha})$ must be an affine function of $\boldsymbol{\alpha}$, such that both $\varphi(\cdot)$ and the coherent state overlaps of the form $\braket{f(\boldsymbol{\chi}(\boldsymbol{\beta}))}{f(\boldsymbol{\chi}(\boldsymbol{\alpha}))}$ remain Gaussian functions of the integration variables.
\end{enumerate}

Condition (1) is satisfied whenever the initial driving field $\ket{\Phi(t_0)}$ is a Gaussian state, whereas condition (2) depends on the way in which the HHG process develops.~For monochromatic driving fields, the HHG spectral amplitudes $\chi(\boldsymbol{\alpha})$ generally depend nonlinearly on the driving field amplitude~\cite{kulander_theory_1990,lhuillier_calculations_1992,weissenbilder_how_2022}; yet, this nonlinearity does not necessarily lead to observable non-Gaussian harmonic states for bright coherent drivers~\cite{lewenstein_generation_2021,stammer_theory_2025}, whose phase-space distributions are narrowly localized. However, for drivers with large field-amplitude fluctuations, such as BSV, the state samples the nonlinear dependence of $\chi(\boldsymbol{\alpha})$ and condition (2) is no longer satisfied.~Conversely, when considering HHG driven by bichromatic laser fields composed by a strong $\omega$ field of intensity $I_\omega$, assisted by a perturbative field at $2\omega$ of field strength $E_{2\omega}$, the amplitudes of the even and odd harmonic orders are found, up to a phase factor, to scale as~\cite{dahlstrom_quantum_2011,pedatzur_attosecond_2015}
\begin{align}
	&E^{(\text{odd})}
		\propto \sqrt{I_\omega} \cos(\epsilon) \simeq \sqrt{I_\omega},\label{Eq:field:amp:odd}
	\\&E^{(\text{even})}
		\propto \sqrt{I_\omega} \sin(\epsilon) \simeq \epsilon\sqrt{I_\omega},\label{Eq:field:amp:even}
\end{align}
where $\epsilon \propto E_{2\omega}/(2\omega)$.~Consequently, for $E_{2\omega}/(2\omega) \ll 1$, the harmonic amplitudes exhibit an approximately affine dependence on the perturbative driving field, thereby satisfying condition (2).

However, we emphasize, that when $c(\boldsymbol{\alpha})$ is sufficiently localized in phase-space such that $\boldsymbol{\chi}(\boldsymbol{\alpha})$ varies negligibly over its support, the state in Eq.~\eqref{Eq:QO:HHG:gen} can be approximated, after normalization and up to an irrelevant global phase, as
\begin{equation}
	\ket{\Phi(t)}
		= \ket{\boldsymbol{\alpha}_0 + \boldsymbol{\chi}_{\text{dr}}(\boldsymbol{\alpha}_0)}
			\bigotimes_{q\in \text{HH}} \ket{\chi_q(\boldsymbol{\alpha}_0)},
\end{equation}
where $\boldsymbol{\alpha}_0$ denotes the center of the Gaussian distribution associated with $c(\boldsymbol{\alpha})$.~This approximation holds irrespective of the single- or multimode nature of the driving field, and therefore provides a sufficient condition for the Gaussianity of the generated state, corresponding to a separable multimode coherent state.~Nonetheless, when the harmonic spectral amplitudes vary significantly over the support of $c(\boldsymbol{\alpha})$, bichromatic driving fields could give rise to non-trivial multimode Gaussian states.~This occurs, for instance, for bright squeezed driving fields, which we analyze explicitly in the following section.

It is important to emphasize that, in this work, we use the strong-field approximation (SFA)~\cite{lewenstein_theory_1994,amini_symphony_2019,RBSFA} to evaluate the harmonic intensities and their dependence on the $2\omega$ field strength. However, we expect the main conclusion relevant to the present analysis, namely the linear dependence of the harmonic amplitudes on the perturbative $2\omega$ field, to be qualitatively robust beyond the SFA. This leading-order dependence follows from the perturbative expansion in the weak $2\omega$ field together with the symmetry-imposed selection rules of the bichromatic driving configuration, which determines the powers of $E_{2\omega}$ allowed for a given harmonic order (see Eqs.~(8) and (9)). Consequently, theoretical descriptions that preserve the relevant symmetry, including Coulomb-corrected  or multi-electron approaches, are expected to exhibit the same leading-order power dependence~\cite{pedatzur_attosecond_2015,tzur_measuring_2025}. A similar argument applies to solid-state HHG provided that the corresponding crystal symmetry and driving configuration lead to the same selection rules~\cite{stammer_symmetry_2026}. Nevertheless, we acknowledge that the precise value of the linear coefficient, as well as the higher-order corrections that determine the range of validity of the perturbative regime, can depend on the specific HHG model.

\subsection{HHG driven by bright squeezed states}\label{Sec:ThBck:BSV}
Recent advances in high-gain spontaneous parametric down-conversion have enabled the generation of BSV sources~\cite{spasibko_multiphoton_2017,manceau_indefinite-mean_2019} with sufficient intensity to induce strong-field ionization in different materials~\cite{heimerl_multiphoton_2024,heimerl_driving_2025,lyu_attosecond_2026} and, importantly, to drive HHG in both monochromatic~\cite{rasputnyi_high_2024,lyu_attosecond_2026} and bichromatic configurations~\cite{lemieux_photon_2025,tzur_attosecond-resolved_2025}, either independently or assisting a strong coherent pump.~To account for both situations with a unified framework, we consider the driving field initially prepared in the state $\ket{\Phi(t_0)} = \ket{\alpha_1} \otimes \hat{S}_2(r)\ket{0_2}$, where $\hat{S}_2(r) = \text{exp}[r(\hat{a}_2^2 - \hat{a}_2^{\dagger 2})/2]$ is the squeezing operator and, for simplicity, we restrict the discussion to real-valued squeezing strengths $r$.~In terms of the coherent state expansion introduced in Eq.~\eqref{Eq:init:state:coh:expansion}, the state can be written as
\begin{equation}
	\ket{\Phi(t_0)}
		= \int \dd^2 \alpha\ c_{\text{BSV}}(\alpha)
			 \ket{\text{g}}\otimes
			 	\ket{\alpha_1}\otimes
			 		 \ket{\alpha}
			 		 	\bigotimes_{q\in\text{HH}} \ket{0_q},
\end{equation}
where, unlike in Eq.~\eqref{Eq:init:state:coh:expansion}, only the squeezed $2\omega$ contribution is expanded in the coherent state basis, since the $\omega$ field already remains in a coherent state.~The corresponding expansion coefficients are given by
\begin{equation}
	\begin{aligned}
	c_{\text{BSV}}(\alpha)
		&= \dfrac{e^{-\frac{\alpha_x^2}{e^{r}\cosh(r)} 
				-\frac{\alpha_y^2}{e^{-r}\cosh(r)}
				+i 2\alpha_x\alpha_y \tanh(r)}}{\pi \sqrt{\cosh(r)}},
	\end{aligned}
\end{equation} 
where $\alpha = \alpha_x + i \alpha_y$.~Importantly, $c_{\text{BSV}}(\alpha)$ is a complex Gaussian function of $\alpha$ and, consequently, since the quantum optical state after HHG reads 
\begin{equation}\label{Eq:HHG:BSV:gen}
	\begin{aligned}
	\ket{\Phi(t)}
		&= \int \dd^2 \alpha\ c_{\text{BSV}}(\alpha)
				\bigg[
					e^{i\varphi_1(\alpha)}
					\bigotimes_{q\in \text{odd}}\!\!
					\ket{\alpha_1 \delta_{q,1} + \chi_q(\alpha)}\!
				\bigg]
		\\&\hspace{1.5cm}\times\!
					\bigg[
						e^{i\varphi_2(\alpha)}
						\bigotimes_{q\in \text{even}}\!\!
						\ket{\alpha \delta_{q,2} + \chi_q(\alpha)}\!
					\bigg],
	\end{aligned}
\end{equation}
its Gaussianity is entirely determined by the functional dependence of $\chi_q(\alpha)$ on the integration variable.

\begin{figure}
	\centering
	\includegraphics[width=1\columnwidth]{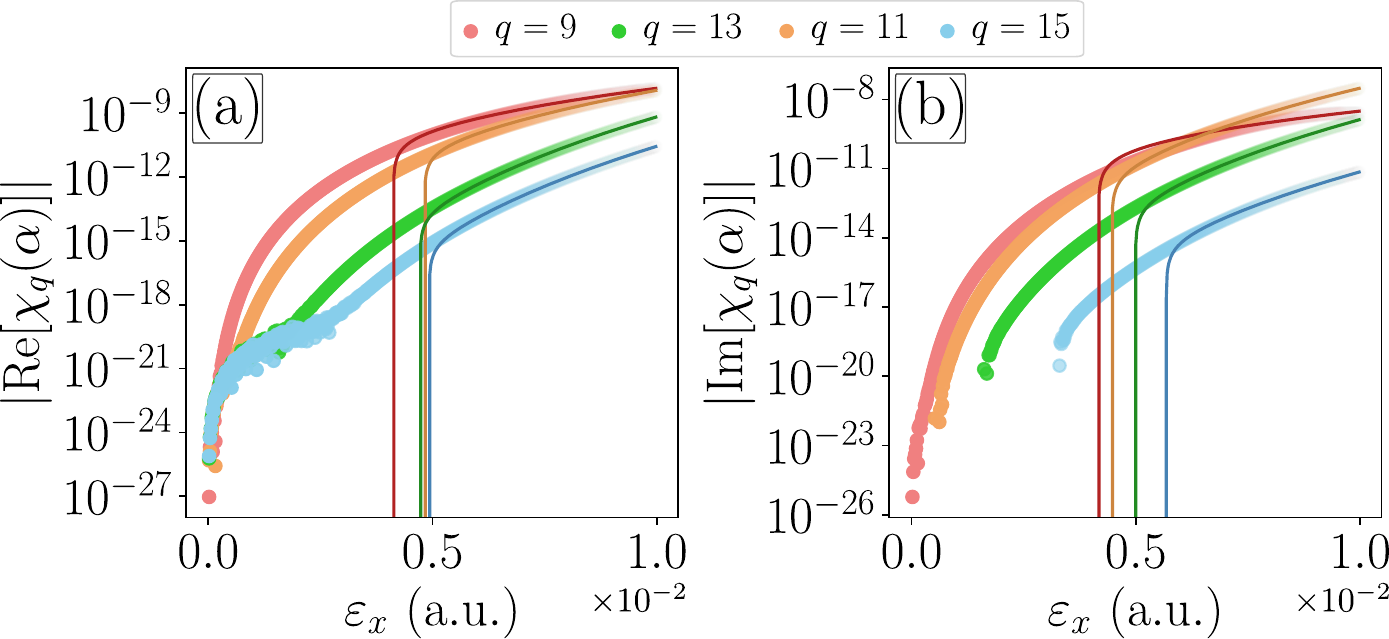}
	\caption{Dependence of the (a) real and (b) imaginary parts of the HHG spectral amplitudes for a BSV driver as a function of the electric field strength $\varepsilon_x = 2\kappa \alpha_x$.~The transparency of each point highlights the probability of finding such field component within the BSV field distribution.~The continuous curves show fits of the form $\chi_q(\alpha) = A_{q} \alpha_x^{p_{q,x}} + i B_q \alpha_y^{p_{q,x}}$ for each harmonic mode.~Calculations have been carried performed for hydrogen ($I_p \simeq 0.5$ a.u.), using a driving field of frequency $\omega = 0.057$ a.u., and a squeezing intensity $I_{\text{squ}} = \kappa^2 \sinh[2](r) = 2.5 \times 10^{-6}$ a.u.~Here, $q$ represents harmonic orders of the $2\omega$ drive.}
	\label{Fig:BSV:fits}
\end{figure}

For a monochromatic $2\omega$ driving field consisting solely of a BSV source ($\alpha_1 = 0$) with sufficient intensity to independently drive HHG $(r \gtrsim 15)$, we find $\chi_q(\alpha) = 0$ for the $4q$ harmonic orders, corresponding to the even harmonic orders of the $2\omega$ squeezed drive.~In contrast, for the $2(2q+1)$ harmonic orders, namely the odd harmonics of the $2\omega$ drive, we approximately obtain the scaling $\chi_q(\alpha) = A_{q} \alpha_x^{p_{q,x}} + i B_q \alpha_y^{p_{q,y}}$ [Fig.~\ref{Fig:BSV:fits}] where, for all generated harmonic orders, the fitted exponents satisfy $p_{q,x},p_{q,y} \gtrsim 10$, highlighting the strong non-linear harmonic response.~Consequently, the harmonic response depends highly nonlinearly on the coherent state amplitude $\alpha$, strongly distorting the initial Gaussian structure of the BSV distribution after the HHG process and thereby generating a highly non-Gaussian quantum optical state.~As a result, the harmonic modes exhibit strong super-Poissonian photon-number statistics with $g^{(2)}\gg 1$, as recently reported~\cite{petrovic_generation_2026,lyu_attosecond_2026,stammer_fluctuation-induced_2026}.~We note that this nonlinear response is dominated by the antisqueezed quadrature $\varepsilon_x = 2\kappa \alpha_x$ whenever $r>0$.~Along the squeezed quadrature, $\varepsilon_y = 2 \kappa\alpha_y$, the distribution $c_{\text{BSV}}(\alpha)$ becomes extremely narrow compared to the scale over which $\chi_q(\alpha)$ significantly varies with $\alpha_y$, such that one can safely approximate $\chi_q(\alpha) \simeq \chi_q(\alpha_x)$.

\begin{figure}
	\centering
	\includegraphics[width=1\columnwidth]{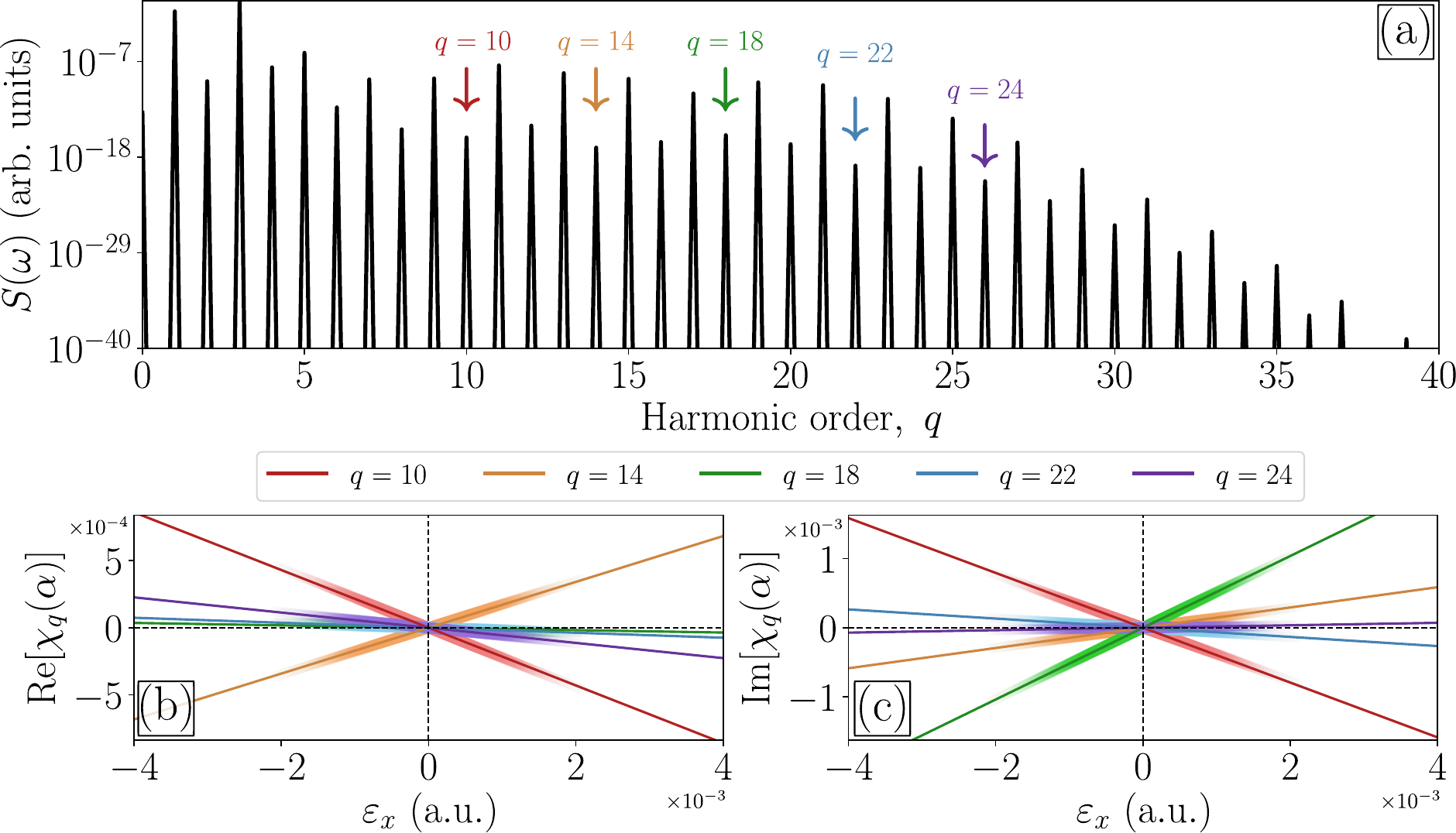}
	\caption{(a) HHG spectrum and (b,c) dependence of the real and imaginary parts, respectively, of the HHG spectral amplitudes for a bichromatic configuration, consisting of a strong coherent field at frequency $\omega$ and a perturbative BSV component at frequency $2\omega$.~The transparency of each point~in panels (b) and (c) highlights the probability of finding such field component within the BSV field distribution.~The continuous curves in panels (b) and (c) display the linear fits of the form $\chi_q(\alpha) = (A_{q} + iB_q) \alpha_x$ for each harmonic mode.~Calculations have been performed for xenon ($I_p \simeq 0.446$ a.u.), using a driving field frequency $\omega = 0.057$ a.u., a coherent field strength $\varepsilon_{\omega} = 0.053$ a.u., and a squeezing intensity for the $2\omega$ component $I_{\text{squ}} = \kappa^2 \sinh[2](r) = 10^{-7}$ a.u.}
	\label{Fig:Bch:fits}
\end{figure}

The situation changes when considering a bichromatic driving field composed of a strong coherent component at frequency $\omega$, with $\abs{\alpha_1} \gg 1$ and sufficient to induce HHG on its own, together with a perturbative BSV field at frequency $2\omega$ with intensity enough to perturb the process ($r \sim 10$).~In the absence of the BSV component, the half-cycle inversion symmetry of the HHG process leads exclusively to the generation of odd harmonic orders.~The addition of the weak $2\omega$ field breaks this symmetry, thereby enabling the generation of even harmonic orders~\cite{dahlstrom_quantum_2011,pedatzur_attosecond_2015}, whose intensities remain several orders of magnitude smaller than those of the odd harmonics under perturbative conditions [see Fig.~\ref{Fig:Bch:fits}~(a) for an example where the HHG spectra $S(\omega_q) = \langle \Phi(t)\vert \hat{a}^\dagger_q\hat{a}_q\vert \Phi(t)\rangle$ is shown].~Within this regime, we find that the harmonic amplitudes associated with the odd orders become effectively independent of the perturbative field, i.e., $\chi_{2q+1}(\alpha) = \chi_{2q+1}$, whereas the even harmonic amplitudes exhibit an approximately linear dependence of the form
\begin{equation}\label{Eq:chi:even:dep}
	\chi_{2q}(\alpha) 
		= A_{2q} \alpha_x, \quad A_{2q} \in \mathbbm{C},
\end{equation}
as shown in Fig.~\ref{Fig:Bch:fits}~(b) and (c), and anticipated from the scaling relations in Eqs.~\eqref{Eq:field:amp:odd} and \eqref{Eq:field:amp:even}.~Consequently, both conditions (1) and (2) are satisfied, implying that the resulting quantum optical state after HHG possesses a Gaussian structure and factorizes into uncorrelated even and odd channels,
\begin{equation}\label{Eq:separable}
	\ket{\Phi(t)}
		= \ket{\Phi_{\text{odd}}(t)}
				\otimes \ket{\Phi_{\text{even}}(t)}.
\end{equation}
The odd sector is given by
\begin{equation}
	\ket{\Phi_{\text{odd}}(t)} 
		= e^{i\varphi_1}\bigotimes_{q=0}
			\ket{\alpha_{2q+1} \delta_{2q+1,1} + \chi_{2q+1}},
\end{equation}
corresponding to a separable multimode coherent state. In contrast, the even sector takes the form 
\begin{equation}\label{Eq:BSV:even}
	\begin{aligned}
	\ket{\Phi_{\text{even}}(t)}
		&= \int \dd^2 \alpha \
				c_{\text{BSV}}(\alpha)
					e^{i\varphi_2(\alpha)}
						\\& \hspace{1.5cm}
						\bigotimes_{q=1}
						\ket{\alpha \delta_{2q,2} + A_{2q} \alpha_x},
	\end{aligned}
\end{equation}
which generally corresponds to a non-trivial multimode Gaussian state.~Importantly, the validity of Eq.~\eqref{Eq:chi:even:dep}, and therefore of Eq.~\eqref{Eq:BSV:even}, depends critically on the squeezing strength.~As the squeezing parameter $r$ increases, the linear approximations in Eqs.~\eqref{Eq:field:amp:odd} and \eqref{Eq:field:amp:even} progressively break down, leading to nonlinear dependencies of the harmonic amplitudes on the driving field and, therefore, to the emergence of non-Gaussian features in the emitted radiation.

\section{RESULTS}
Having identified bichromatic schemes as a route for generating non-trivial multimode Gaussian states of light via HHG, in this section we characterize their properties.~To this end, we introduce a single-quadrature expansion of BSV light that enables a fully analytical description of the resulting state.~We then analyze its single- and multimode properties, as well as the quantum correlations it exhibits, and benchmark their utility within the paradigmatic quantum teleportation protocol.~Finally, we investigate the conditions under which Gaussianity breaks down in this bichromatic configuration, comparing it with recent experimental results~\cite{tzur_attosecond-resolved_2025}.

\subsection{Single-quadrature expansion of the BSV driver}
One of the defining features of squeezed states is the anisotropy of their fluctuations along conjugate optical quadratures, namely $\hat{X}_1 = (\hat{a} + \hat{a}^\dagger)/\sqrt{2}$ and $\hat{X}_2 = i(\hat{a}^\dagger - \hat{a})/\sqrt{2}$.~For a single-mode squeezed state, these fluctuations are given by $(\Delta X_1)^2 = e^{2r}/2$ and $(\Delta X_2)^2 = e^{-2r}/2$, such that the Heisenberg uncertainty relation is saturated for all values of the squeezing parameter $r$, i.e., $(\Delta X_1)^2 (\Delta X_2)^2 = 1/4$.~In the regime relevant to this work, namely $r \sim 10$, one has $e^{\pm 2r} \sim 10^{\pm8}$, resulting in an extremely elongated phase-space distribution along one quadrature and a highly localized one along its conjugate counterpart.~Motivated by this pronounced anisotropy, we introduce a simplified one-quadrature expansion for the BSV field which, for squeezing parameters $r >(1/2)\ln(3/2)\simeq 0.203$, well within our parameter regime, exactly captures the dominant statistical properties of the state (see Supplementary Material~\ref{Sec:SM:1D:model}), while allowing for a fully analytical treatment of the quantum optical properties of the HHG state.~Specifically, we consider the single-quadrature expansion
\begin{equation}\label{Eq:One:quad:model}
	\relaxket{\Tilde{\Phi}_2(t_0)}
	= \int_{\mathbbm{R}} \dd \alpha \ \Tilde{c}(\alpha)\ket{\alpha},
\end{equation}
with the probability amplitude defined as
\begin{equation}
	\tilde{c}(\alpha)
	= \sqrt{\dfrac{(1+\sigma)^{\frac12}}{\pi \sigma}} \exp[-\dfrac{\alpha^2}{\sigma}],
	\ \text{with}\ \sigma = e^{\bar{r}} \cosh(\bar{r}).
\end{equation}

\begin{figure}
	\centering
	\includegraphics[width=1\columnwidth]{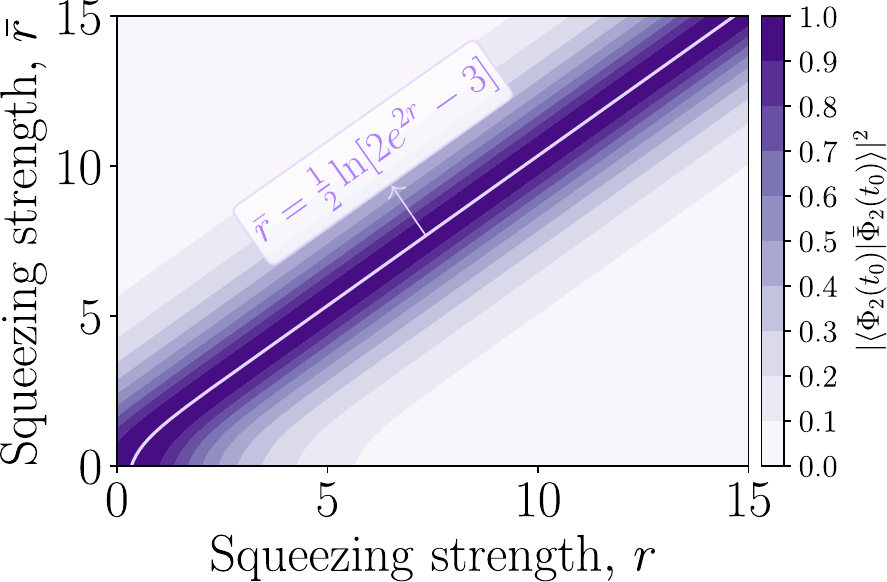}
	\caption{Fidelity between the single-quadrature expansion in Eq.~\eqref{Eq:One:quad:model} and a squeezed state $\hat{S}(r)\ket{0}$ as a function of $r$ and $\bar{r}$. The light curve highlights the $\bar{r} = \ln[2e^{2r}-3]/2$ values, along which $\mathcal{F}(r,\bar{r}) = 1$, as analytically demonstrated in the Supplementary Material~\ref{Sec:SM:1D:model}.}
	\label{Fig:Fid}
\end{figure}

For this family of states, the quadrature fluctuations become $(\Delta X_1)^2 = (1+e^{\bar{r}}\cosh(\bar{r}))/2$ and $(\Delta X_1)^2 = [2(1+e^{\bar{r}}\cosh(\bar{r}))]^{-1}$, such that ($\Delta X_1)^2 > 1/2$ and $(\Delta X_2)^2 < 1/2$, which coincide with the ones presented earlier for a squeezed state defined through the squeezing operator $\hat{S}(r)$ when choosing $\bar{r} = \ln[2e^{2r}-3]/2$.~To quantify more precisely the similarity between both states, we compute the fidelity with respect to $\ket{\Phi_2(t_0)} = \hat{S}_2(r)\ket{0_2}$, obtaining
\begin{equation}\label{Eq:Fid:1D}
	\begin{aligned}
		\mathcal{F}(r,\bar{r}) 
		&= \big\lvert\!\langle \Phi_2(t_0)\vert \Tilde{\Phi}_2(t_0)\rangle\!\big\rvert^2
		\\&
		= 
		\sqrt{\dfrac{4(1+\sigma)^{\frac12}}{\sigma\cosh(r)[2+4 e^{-\bar{r}}\sech(\bar{r})-2\tanh(r)]}},
	\end{aligned}
\end{equation}
which for the aforementioned relation between $r$ and $\bar{r}$, yields $\mathcal{F}(r,\bar{r}) = 1$, as shown in Fig.~\ref{Fig:Fid}, and analytically demonstrated in Supplementary Material~\ref{Sec:SM:1D:model}.~We note, however, that for $\bar{r} < 0$ the overlap quickly decays to zero as the squeezing orientations defined by $\hat{S}(r)\ket{0}$ and Eq.~\eqref{Eq:One:quad:model} are orthogonal, in which case the integration contour in the aforementioned definition has to be properly adjusted.~Finally, for all values of $\bar{r}$ the state preserves the even photon number parity  characteristic of squeezed states, i.e.,
\begin{equation}
	\begin{aligned}
		P(n) &=  \big\lvert\langle n\vert \Tilde{\Phi}_2(t_0)\rangle\big\rvert^2
		\\&= \left\{
		\begin{aligned}
			& 0, \ \text{if} \ n \ \text{is odd}\\
			& \dfrac{[(n-1)!!]^2 \sigma^n \sqrt{1+\sigma}}{(2+\sigma)^{n+1}n!},\ \text{if} \ n \ \text{is even}
		\end{aligned}
		\right. .
	\end{aligned}
\end{equation}

\subsection{Effective single-mode structure of the post-HHG state}
Within the perturbative bichromatic setting identified in Sec.~\ref{Sec:ThBck:BSV}, where the harmonic amplitudes depend linearly on the squeezed driving field, the quantum optical state associated with the even harmonic orders acquires a particularly simple structure.~Using the single-quadrature expansion introduced earlier, the quantum optical state of the even harmonic orders after HHG can be written as
\begin{equation}\label{Eq:HHG:approx:BSV}
	\relaxket{\Tilde{\Phi}_{\text{even}}(t)}
			= \int_{\mathbbm{R}} \dd \alpha \ \tilde{c}(\alpha)
					\bigotimes_{q=1} \hat{D}_q\big( [\delta_{2q,2}+A_{2q}] \alpha\big)
						\ket{\boldsymbol{0}},
\end{equation}
where, in what follows, we set $\varphi_2(\alpha) = 0$ as this choice yields the best agreement with recent experimental observations discussed in Sec.~\ref{Sec:Brkdwn:gaussian}. A detailed analysis of the influence of this phase on the quantum optical properties of the generated state is provided in the Supplementary Material~\ref{Sec:SM:Phase:influence}.

A key feature of Eq.~\eqref{Eq:HHG:approx:BSV} is that the multimode displacement structure can be recast into the excitation of a single collective bosonic mode.~Indeed, due to the commutation relation $[\hat{D}_q(\cdot), \hat{D}_{q'}(\cdot)] = 0$ for $q\neq q'$, together with the linear dependence of the harmonic amplitudes on $\alpha$, the tensor product displacement operator becomes
\begin{equation}
	\begin{aligned}
	&\bigotimes_{q=1} \hat{D}_q\big([\delta_{2q,2}+A_{2q}] \alpha\big)
		\\&\hspace{1.5cm} = \exp[\alpha \!\sum_{q=1} (\delta_{2q,2}+A_{2q})\hat{a}^\dagger_{2q} - \text{h.c.}].
	\end{aligned}
\end{equation}
This naturally motivates the introduction of the collective HHG excitation mode~\cite{lewenstein_generation_2021,stammer_theory_2022,stammer_high_2022}
\begin{equation}\label{Eq:global:mode:def}
	\hat{A} 
		= \dfrac{1}{\bar{\chi}}
			\sum_{q=1} A_{2q} \hat{a}_{2q},
	 \ \text{with} \
	 	 \bar{\chi}^2 = \sum_{q=1} \abs{A_{2q}}^2,
\end{equation}
which satisfies the bosonic commutation relation $[\hat{A},\hat{A}^\dagger] = \mathbbm{1}$, in terms of which Eq.~\eqref{Eq:HHG:approx:BSV} can be rewritten as
\begin{equation}
	\relaxket{\Tilde{\Phi}_{\text{even}}(t)}
		= \sqrt{\dfrac{(1+\bar{\chi}^2\sigma)^{\frac12}}{\pi \sigma}}
			\! \int_{\mathbbm{R}}\!\! \dd \alpha \ \tilde{c}(\alpha)
				\hat{D}_{\text{even}}(\bar\chi \alpha) \ket{0} ,	
\end{equation}
where $\hat{D}_{\text{even}}(\alpha) = \text{exp}[\alpha \hat{A}^\dagger-\text{h.c.}]$ denotes the displacement operator associated with this collective excitation, with $\ket{0}$ representing its vacuum state.

Using this collective mode representation, the Gaussian character of the generated state becomes manifest. In particular, its wavefunction in an arbitrary quadrature basis $\hat{X}_{\theta} = \hat{X}_1 \cos(\theta) + \hat{X}_2\sin(\theta)$ can be obtained analytically as
\begin{equation}\label{Eq:QO:wavefunction}
	\begin{aligned}
	\Tilde{\Phi}_{\text{even}}(X_{\theta})
		&= \bra{X_1}\! e^{i\theta \hat{A}^\dagger \hat{A}}\relaxket{\Tilde{\Phi}_{\text{even}}(t)}
		\\&
		= \sqrt{\dfrac{(1+\bar{\chi}^2\sigma)^{\frac12}}{\pi \sigma}}
				\exp[-\frac{C(\theta) X_1^2}{2}]
	\end{aligned}
\end{equation}
where $C(\theta)$ is the complex-valued function
\begin{equation}
	C(\theta)
		= \dfrac{\bar{\chi}^2\sigma \cos(\theta)\sin(\theta) + i [1+\bar{\chi}^2\sigma\sin^2(\theta)]}{-\bar{\chi}^2\sigma \cos(\theta)\sin(\theta) + i [1+\bar{\chi}^2\sigma\cos^2(\theta)]}.
\end{equation}
Since $\text{Re}[C(\theta)]\geq 0$ for all $\theta$, the quadrature wavefunction remains Gaussian for any quadrature basis, thereby explicitly confirming the Gaussian nature of the generated state.~Consequently, the state can be fully characterized through its covariance matrix $\Gamma$, with elements given by
\begin{equation}
	\Gamma_{i,j}
		= \dfrac12
			\big[
				\langle \hat{X}_i \hat{X}_j \rangle
				+ \langle \hat{X}_j \hat{X}_i \rangle
			\big],
\end{equation}
where the first moments vanish, i.e., $\langle \hat{X}_i\rangle = 0$, due to the vacuum centered nature of the state. This yields
\begin{equation}\label{Eq:SM:Cov:matrix}
	\Gamma 
		= \mqty(\dfrac12 [1+\bar{\chi}^2\sigma] & 0 \\
						0 & \dfrac{1}{2(1+\bar{\chi}^2\sigma)}),
\end{equation}
from which we observe that, for $\bar{\chi}=1$, the quadrature variances reduce to those of the effective BSV state prior to the strong-field interaction.~Furthermore, although the analysis presented in Sec.~\ref{Sec:ThBck} was performed within the single-active-electron approximation, the contribution of $N$ atoms emitting independently and coherently can be effectively incorporated through the substitution $\chi_q(\boldsymbol{\alpha}) \to N \chi_q(\boldsymbol{\alpha})$~\cite{lewenstein_generation_2021,stammer_quantum_2023}, which naturally reproduces the $N^2$ scaling of the harmonic intensity~\cite{eberly_spectrum_1992,rivera-dean_quantum-optical_2024,stammer_theory_2025}.~Since $\chi_q(t) \propto \kappa$, we introduce the effective collective coupling parameter $\varrho = N \kappa$ which we use as the relevant dynamical parameter throughout the following analysis. Simultaneously, we restrict to values of $\varrho$ that satisfy the condition $\bar{\chi}\leq 1$, thereby ensuring that the emitted harmonic field remains smaller than the driving field amplitude.

\begin{figure}
	\centering
	\includegraphics[width=1\columnwidth]{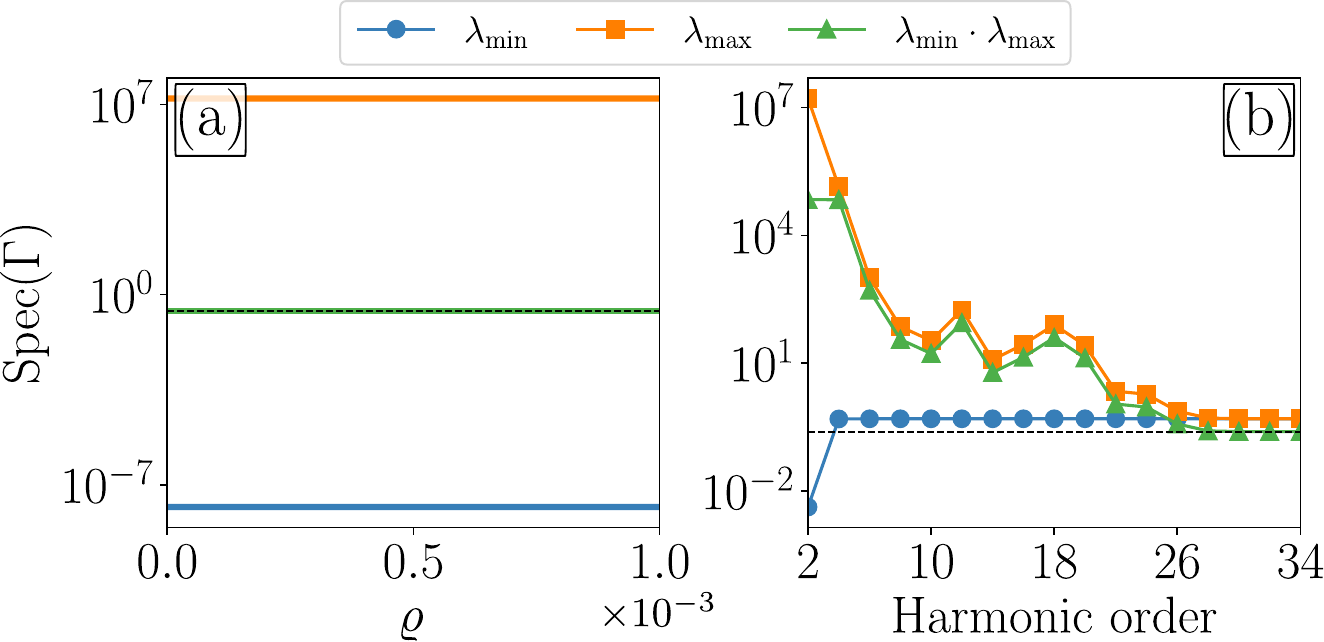}
	\caption{Eigenvalues of the covariance matrix for (a) the collective mode defined in Eq.~\eqref{Eq:global:mode:def}, and (b) individual harmonic modes obtained after tracing out all other modes, for $\varrho = 10^{-3}$.~The black dashed line indicates the Heisenberg undecrtainty limit $\lambda_{\text{min}}\cdot \lambda_{\text{max}} = 1/4$.~Calculations have been done for xenon atoms ($I_p\simeq 0.446$ a.u.), using a driving field frequency $\omega = 0.057$ a.u., a coherent field strength $\varepsilon_{\omega} = 0.053$ a.u., and a squeezing parameter $\bar{r} = 9$ $(r \simeq 8.65)$ corresponding to a squeezing intensity $I_{\text{squ}} \sim 10^{-7}$ a.u.~for $\kappa = 5 \times 10^{-8}$ a.u.}
	\label{Fig:Eigenvalues:covariance}
\end{figure}

Figure~\ref{Fig:Eigenvalues:covariance}~(a) displays the eigenvalues of Eq.~\eqref{Eq:SM:Cov:matrix} as a function of the effective coupling pameter $\varrho$, enforcing the conditions discussed above.~As observed, the global collective mode preserves a squeezed Gaussian structure throughout the considered parameter regime, with covariance eigenvalues reaching $\lambda_{\text{max}} \sim 10^{7}$ and $\lambda_{\text{min}} \sim 10^{-7}$, while preserving its minimum-uncertainty structure, i.e., $\lambda_{\text{max}}\lambda_{\text{min}} = \det[\Gamma] = 1/4$.~The situation changes, however, when considering reduced subsets of harmonic modes.~In analogy with the collective excitation mode introduced in Eq.~\eqref{Eq:global:mode:def}, we partition the even part of the harmonic spectrum in two complementary subsets $\mathsf{A}$ and $\mathsf{B}$, satisfying $\mathsf{A}\cup \mathsf{B} = \{q = 2n: n\in \mathbbm{N}\}$, and define the corresponding collective operators
\begin{equation}\label{Eq:bipartitions}
	\hat{A} 
		= \dfrac{1}{\bar{\chi}_{\mathsf{A}}}
				\sum_{q\in \mathsf{A}} A_q \hat{a}_q
	\quad \text{and} \quad
	\hat{B} 
		= \dfrac{1}{\bar{\chi}_{\mathsf{B}}}
				\sum_{q\in \mathsf{B}} A_q \hat{a}_q,
\end{equation}
which satisfy $[\hat{A},\hat{B}] = 0$.~The joint quantum optical state can then be written as
\begin{equation}
	\relaxket{\Tilde{\Phi}_{\text{even}}(t)}
		= \sqrt{\dfrac{(1+\bar{\chi}^2\sigma)^{\frac12}}{\pi \sigma}}
				\! \int_{\mathbbm{R}}\!\! \dd \alpha \ \tilde{c}(\alpha)
					\ket{\chi_{\mathsf{A}}\alpha, \chi_{\mathsf{B}}\alpha}.
\end{equation}
The associated covariance matrix takes the block form
\begin{equation}\label{Eq:tot:Gamma}
	\Gamma 
		= \mqty(\Gamma^{(\mathsf{A})} & \Lambda \\
					 \Lambda^T &  \Gamma^{(\mathsf{B})}),
\end{equation}
where $\Gamma^{(\mathsf{A})}$ and $\Gamma^{(\mathsf{B})}$ denote the reduced covariance matrices of each subsystem, effectively obtained from Eq.~\eqref{Eq:SM:Cov:matrix} through the substitution $\bar{\chi} \to \bar{\chi}_{\mathsf{A}/\mathsf{B}}$, while $\Lambda$ contains the cross-correlations between both sectors.

\begin{figure*}
	\centering
	\includegraphics[width=1\textwidth]{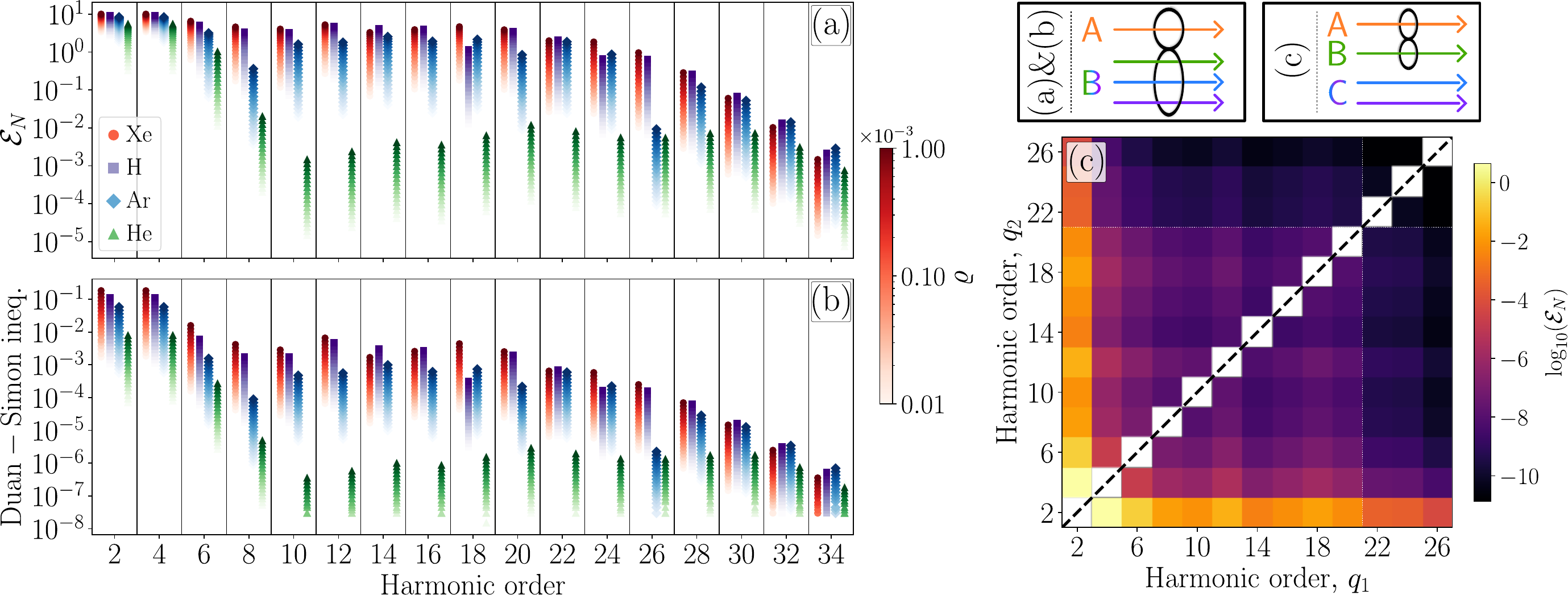}
	\caption{Entanglement characterization of the resulting quantum optical state, with bipartitions chosen as shown in the inset at the top right corner of the figure.~(a),(b) Entanglement between one harmonic mode and all other modes, as a function of the harmonic mode index, characterized through the logarithmic negativity and the Duan-Simon criterion, respectively.~(c) Pairwise entanglement evaluated using the logarithmic negativity for $\varrho = 10^{-3}$.~The same parameters as in Fig.~\ref{Fig:Eigenvalues:covariance} have been used for the field, while different atomic systems have are considered, with ionization potentials $I_p^{(\text{Xe})} \simeq 0.446\ \text{a.u.} <\  I_p^{(\text{H})} \simeq 0.5 \ \text{a.u.} \ <\ I_p^{(\text{Ar})} \simeq 0.579 \ \text{a.u.} \ <\ I_p^{(\text{He})} \simeq 0.903$ a.u.}
	\label{Fig:Ent:DSI}
\end{figure*}

To analyze the properties of individual harmonics, we choose $\mathsf{A} = \{2q\}$, therefore $\mathsf{B} = \bar{\mathsf{A}}$, such that subsystem $\mathsf{A}$ contains a single harmonic mode.~Figure~\ref{Fig:Eigenvalues:covariance}~(b) displays the corresponding covariance eigenvalues as a function of the harmonic order for $\varrho = 10^{-3}$.~Unlike the global collective mode, individual harmonics no longer saturate the minimum-uncertainty condition (green curve with triangular markers), indicating that they correspond to mixed Gaussian states arising from tracing out the correlations with the remaining harmonic modes~\cite{rivera-dean_attosecond_2026}.~As observed, while the global mode structure is indeed compatible with that of a global squeezed-like state, each harmonic mode lies in a mixed state with squeezed quadrature fluctuations saturating the vacuum limit $\lambda_{\text{min}} = 0.5$ (blue curve with circular markers), the notable exception being the fundamental $2\omega$ mode for which $\lambda_{\text{min}}\sim 10^{-2}$.~For harmonic orders below the cutoff region (around the 20th harmonic), the largest covariance eigenvalue satisfies $\lambda_{\text{max}} > 0.5$, exhibiting a structure reminiscent of a conventional HHG spectrum:~an initial decay at low harmonic orders followed by a plateau extending up to the cutoff region, after which the eigenvalues progressively approach the vacuum limit $\lambda_{\text{max}} = \lambda_{\text{min}} = 0.5$.

This behavior indicates that the fundamental mode retains the strongest non-classical fluctuations after the strong-field interaction, while simultaneously developing significant correlations with the generated harmonics.~In the following subsection, we analyze the quantum nature of these correlations in greater detail.

\subsection{Entanglement structure of the generated harmonics}\label{Sec:Ent}

\subsubsection{Entanglement characterization}
Within the continuous variable quantum information framework, the entanglement structure of a Gaussian state is fully encoded in its covariance matrix~\cite{horodecki_quantum_2009}, making its characterization particularly tractable.~For bipartitions of the form in Eq.~\eqref{Eq:bipartitions}, the positive partial transpose (PPT) criterion~\cite{peres_separability_1996,horodecki_separability_1996} provides a necessary and sufficient condition for separability~\cite{simon_peres-horodecki_2000}, and the degree of entanglement can be quantified exactly through the logarithmic negativity~\cite{horodecki_quantum_2009,serafini_entanglement_2023}
\begin{equation}
	\mathcal{E}_N 
		= \max\big\{0,-\log_2(2\tilde{\lambda}_-)\big\}.
\end{equation}
Here, $\tilde{\lambda}_-$ denotes the smallest symplectic eigenvalue of the partially transposed covariance matrix $\Gamma^{T_\mathsf{B}}$, taken with respect to $\mathsf{B}$ without loss of generality, obtained as the smallest absolute eigenvalue of $i \Omega \Gamma^{T_{\mathsf{B}}}$ where $\Omega$ is the symplectic form
\begin{equation}
	\Omega = \mqty(0&1&0&0\\
								-1&0&0&0\\
								0&0&0&1\\
								0&0&-1&0).
\end{equation}
 Thus, the state is entangled if and only if $\Tilde{\lambda}_- < 1/2$, in which case $\mathcal{E}_N >0$ quantifies the amount of bipartite entanglement.

Figure~\ref{Fig:Ent:DSI}~(a) displays the logarithmic negativity obtained when choosing $\mathsf{A} = \{2q\}$ and $\mathsf{B} = \bar{\mathsf{A}}$, as a function of the harmonic order and for different atomic species and values of the effective coupling parameter $\varrho$.~Consistent with the covariance matrix analysis of Fig.~\ref{Fig:Eigenvalues:covariance}~(b), the strongest entanglement is found between the second harmonic mode and the remaining ones.~As the harmonic order increases, the logarithmic negativity initially decreases, subsequently develops a plateau structure, and finally drops abruptly beyond the cutoff region, a behavior directly mirroring the spectral structure of the HHG emission.~Increasing $\varrho$ systematically enhances the entanglement across all harmonic orders, as a stronger coupling more efficiently transfers the non-classical properties of the BSV into the generated harmonics.~The atomic species, whose ionization potentials satisfy $I^{(\text{Xe})}_{p} <I^{(\text{H})}_{p} <I^{(\text{Ar})}_{p} <  I^{(\text{He})}_{p}$, primarily govern the harmonic cutoff and, therefore, the range over which non-classical correlations persist.~Species with larger ionization potentials sustain entanglement over a broader range of harmonic orders, consistent with their higher cutoff energies.~However, this extension comes at the expense of reduced entanglement magnitude:~atoms with larger ionization potentials are less efficiently ionized by the driving field, resulting in weaker harmonic spectral amplitudes and, therefore, a less efficient transfer of non-classical correlations from the BSV field to the harmonic modes.

From a practical perspective, evaluating the logarithmic negativity requires reconstructing the full covariance matrix of the state. It is therefore relevant to ask whether entanglement can still be witnessed using only a reduced subset of the covariance matrix elements.~In this context, the Duan-Simon criterion~\cite{duan_inseparability_2000,simon_peres-horodecki_2000} becomes particularly useful, providing a sufficient (although not necessary) condition for bipartite entanglement.~Specifically, a Gaussian state is entangled whenever, for some $a \in \mathbbm{R}$,
\begin{equation}
	[\Delta U]^2 + [\Delta V]^2 < a^2 + \dfrac{1}{a^2},
\end{equation}
where $\hat{U} = \abs{a}\hat{X}_{1,\mathsf{A}} + a^{-1} \hat{X}_{1,\mathsf{B}}$ and $\hat{V} = \abs{a}\hat{X}_{2,\mathsf{A}} - a^{-1} \hat{X}_{2,\mathsf{B}}$~\cite{duan_inseparability_2000}.~Expressing the left hand side in terms of the covariance matrix elements, we define the degree of Duan-Simon inequality violation as
\begin{equation}\label{Eq:DSI}
	\text{DSI} 
		= a^2 + \dfrac{1}{s^2} 
			-\sum_{i=1}^2
				\Big[
					\Gamma^{(\mathsf{A})}_{i,i} 
					 + \Gamma^{(\mathsf{B})}_{i,i} 
					+(-1)^{i+1} \Lambda_{i,i},
				\Big].
\end{equation}
such that $\text{DSI} > 0$ constitutes a sufficient condition for entanglement.

Figure~\ref{Fig:Ent:DSI}~(b) displays the DSI under the same conditions used for the logarithmic negativity.~In contrast to the latter, however, the DSI requires optimizing the inequality with respect to the parameter $a$ for each harmonic order and value of $\varrho$.~Therefore, each point in the figure corresponds to a distinct optimized inequality of the form given in Eq.~\eqref{Eq:DSI} (see Supplementary Material \ref{Sec:SM:DSI:opt} for details about the optimization).~Structurally, the DSI follows closely the behavior of both the logarithmic negativity~[Fig.~\ref{Fig:Ent:DSI}~(a)] and the HHG spectrum~[Fig.~\ref{Fig:Bch:fits}~(a)], exhibiting its strongest violations for the lowest harmonic orders.~However, compared to the logarithmic negativity, it becomes noticeable less sensitive to entanglement within the plateau region, where the violation decreases by nearly two orders of magnitude, whereas the logarithmic negativity decreases by less than one order of magnitude.

Finally, we conclude this entanglement characterization subsection by studying pairwise harmonic correlations.~All analyses discussed thus far have focused on the entanglement between a single mode and the remaining harmonic manifold which, although theoretically well defined, may require experimentally demanding measurement schemes.~In particular, reconstructing the covariance matrix of a multimode quantum optical state typically relies on homodyne detection, where the signal field is interfered with a local oscillator engineered to populate the same optical mode and selectively probe specific quadrature components~\cite{roslund_wavelength-multiplexed_2014,medeiros_de_araujo_full_2014}.~For the bipartitions considered in Fig.~\ref{Fig:Ent:DSI}~(a) and (b), this would require constructing local oscillators matching both the individual harmonic mode considered in $\mathsf{A}$ and the collective multimode excitation associated with subsystem $\mathsf{B}$.~Given the broadband spectral nature of the HHG radiation, such collective mode selective measurements become experimentally challenging.

Motivated by this, we instead consider simpler bipartitions involving individual harmonic orders, which can be spectrally resolved by using a prism, for instance, and analyze the amount of entanglement in these.~Specifically, we define the tripartition $\mathsf{A} = \{q_1\}$, $\mathsf{B} = \{q_2\}$ and $\mathsf{C} = \{ q = 2 n, n \in \mathbbm{N}: q \neq q_1,q_2\}$, and study the entanglement between $\mathsf{A}$ and $\mathsf{B}$ after tracing out $\mathsf{C}$.~Figure~\ref{Fig:Ent:DSI}~(c) displays the corresponding logarithmic negativity for $\varrho = 10^{-3}$ and Xe atoms.~Compared to the collective mode analysis, the pairwise harmonic entanglement is significantly reduced throughout the plateau region and becomes practically negligible beyond the cutoff (highlighted by the thin dashed white line).~Nevertheless, consistent with the previous analyses, the strongest correlations are established between the lowest harmonic orders, in particular the $2\omega$ mode, emphasizing its dominant role in the redistribution of non-classical fluctuations across the harmonic spectrum.

However, we highlight that the entanglement properties of the state, as well as its phase-space characteristics, can be affected by macroscopic effects.~While these have been introduced here phenomenologically through the $N$-scaling of the harmonic spectral response, recent studies have found that propagation and phase-matching effects can modify the quantum optical state and its photon statistics~\cite{rivera-dean_propagation_2026,wang_generation_2026}.~Interestingly, Ref.~\cite{wang_generation_2026}, considering a configuration similar to the one studied here but with the roles of the $\omega$ and $2\omega$ fields interchanged, found that different sets of harmonic modes (corresponding in our case to the even and odd sectors) exhibit different optimal gas-jet pressures for phase matching, each within a relatively narrow range. Even under perfect phase matching, however, absorption in the medium can lower the correlations predicted in this section.~We investigate this effect in more detail in Supplementary Material~\ref{Sec:SM:Noise} using a Gaussian loss model.

\subsubsection{Applicability of entanglement to quantum teleportation}\label{Sec:Tel}
While the presence of entanglement across different harmonic orders is fundamentally interesting on its own, the practical relevance of such correlations is ultimately determined by their usefulness in concrete quantum information tasks, namely, by the extent to which they can outperform classical resources.~To benchmark the utility of the generated correlations, we consider continuous variable quantum teleportation~\cite{braunstein_teleportation_1998}, a paradigmatic protocol in Gaussian quantum information~\cite{braunstein_quantum_2005}, for which well-established fidelity bounds exist for both classical and non-classical Gaussian states of light~\cite{adesso_equivalence_2005,mari_optimal_2008}.

The goal of quantum teleporation is to transfer an unknown quantum state $\ket{\psi}$ from one party (Alice) to another (Bob), located at distant positions, using only local operations and classical communication assisted by a shared entangled state $\ket{\Phi}$~\cite{NielsenBookCh1}.~For that purpose, Alice couples the state to be teleported with her share of the entangled resource through a local unitary operation, performs a measurement on the resulting correlated system, and communicates the measurement outcome to Bob through a classical channel.~Conditioned on this information, Bob applies a suitable unitary operation to his share of the entangled state, thereby reconstructing the teleported state at his location.

In the continuous variable setting~\cite{braunstein_teleportation_1998}, the protocol is commonly implemented by interfering the input state $\ket{\psi}$ with Alice's share of the entangled state resource on a balanced beam splitter, followed by quadrature measurements on both beam splitter outputs.~The measurement outcomes determine a displacement operation that Bob applies to his mode, yielding an output state $\hat{\rho}_{\text{out}}$.~Under ideal measurement conditions, the teleportation fidelity $\mathcal{F}_{\text{tel}} = \mel{\psi}{\hat{\rho}_{\text{out}}}{\psi}$ depends directly on the amount of entanglement shared between Alice and Bob.~For two-mode Gaussian states, the teleportation fidelity is bounded as~\cite{mari_optimal_2008}
\begin{equation}\label{Eq:Fid:tel}
	\dfrac{1+2\Tilde{\lambda}_-}{1+6\Tilde{\lambda}_-}
		\leq \mathcal{F}_{\text{tel}} \leq
	\dfrac{1}{1+2\Tilde{\lambda}_-},
\end{equation}
with the upper bound achieved exactly for pure states~\cite{adesso_equivalence_2005} and for mixed states with $\Gamma_{\mathsf{A}} = \Gamma_{\mathsf{B}}$.~Both bounds coincide at $\lambda_- = 1/2$ ($\mathcal{E}_N = 0$), recovering the classical limit $\mathcal{F}_{\text{tel}}^{(\text{cl})} = 1/2$, and perfect teleportation is approached in the limit $\Tilde{\lambda}_- \to 0$.

\begin{figure}
	\centering
	\includegraphics[width=1\columnwidth]{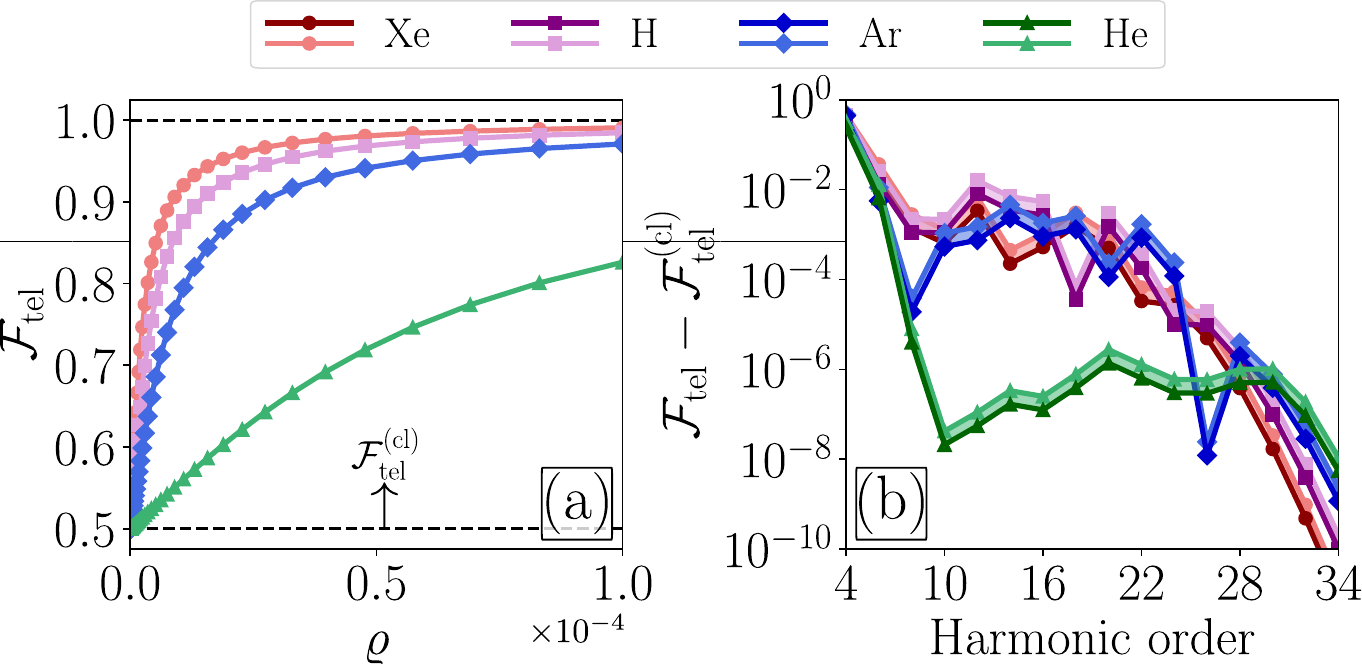}
	\caption{Fidelity of teleporation when (a) using collective mode bipartitions, and (b) using pairwise harmonic correlations.~In panel (b), the lower bounds are shown in dark color while the upper bounds with lighter versions.~The same parameters as in Fig.~\ref{Fig:Ent:DSI} have been used here.}
	\label{Fig:fid:teleport}
\end{figure}

Figure~\ref{Fig:fid:teleport} displays the teleportation fidelity for the two types of bipartitions considered in Fig.~\ref{Fig:Ent:DSI} for the same atomic species.~Panel~(a) shows the case $\mathsf{A} = \{2\}$, $\mathsf{B} = \bar{\mathsf{A}}$, where the entanglement is the greatest~[Fig.~\ref{Fig:Ent:DSI}~(a)] and the resource is a pure two-mode Gaussian state, as a function of $\varrho$.~Panel~(b) shows the case $\mathsf{A} = \{2\}$, $\mathsf{B} = \{q_2\}$, where all other modes are traced out yielding an asymmetric two-mode mixed Gaussian state ($\Gamma_{\mathsf{A}} \neq \Gamma_{\mathsf{B}}$), for which we display the bounds of Eq.~\eqref{Eq:Fid:tel} (light color for the upper bound and dark color for lower bound) as a function of $q_2$ at fixed $\varrho = 10^{-3}$.

In panel (a), the teleportation fidelity grows with $\varrho$ and asymptotically approaches the perfect teleportation scenario, with convergence that is faster for atomic species with lower ionization potential, reflecting the greater efficiency of the HHG process.~In panel (b), only low harmonics orders $q_2$, yield a fidelity appreciably above the classical threshold, reaching values of $\sim\! 0.95$ at $q_2=4$, while modes in the plateau and cutoff region offer a marginal improvement of at most $\sim\! 0.01\%$ over the classical limit.~Intuitively, this contrast arises because the entanglement generated during the HHG process is not confined to any single pair of harmonics, but instead distributed collectively across many harmonic modes.~When only two individual harmonic orders are considered, as in panel (b), most of this entanglement resource is left untapped, since it is shared with the remaining modes. Consequently, the bipartition of panel (a), which groups all of the ``rest'' of the field together against the fundamental, captures the full entanglement available in the state and is consequently able to outperform the classical limit.~Thus, these results suggest that the quantum correlations generated through this bichromatic HHG scheme become useful for quantum information protocols only when the collective mode bipartition $\mathsf{A} = \{2\}$ and $\mathsf{B} = \bar{\mathsf{A}}$ is employed, clearly surpassing the classical bound, whereas the more experimentally accessible pairwise harmonic correlations offer at most a marginal advance over classical strategies.

\subsection{Breakdown of Gaussianity}\label{Sec:Brkdwn:gaussian}
All the analyses performed thus far rely on the validity of Eqs.~\eqref{Eq:field:amp:odd} and \eqref{Eq:field:amp:even}, namely, on the assumption that the harmonic spectral amplitudes scale linearly with $\alpha$.~However, when the squeezing parameter becomes sufficiently large, this approximation breaks down, thereby violating condition (2) introduced in Sec.~\ref{Sec:Conds:Gauss}, with the resulting optical states ceasing to be Gaussian.~In the Supplementary Material~\ref{Sec:SM:Gaussian:valid}, we find that the applicability of the Gaussian approximation remains valid whenever $\gamma \sigma \ll 1$, with $\gamma$ representing the leading higher-order Taylor coefficients in Eqs.~\eqref{Eq:field:amp:odd} and \eqref{Eq:field:amp:even}.

To demonstrate this transition, we evaluate the Wigner function of a given harmonic mode obtained from the general Eq.~\eqref{Eq:QO:HHG:gen} when tailored to our bichromatic setting, defined as~\cite{royer_wigner_1977}
\begin{equation}
	W_q(\beta) 
		= \dfrac{2}{\pi} \mel{\Phi_{\text{even}}(t)}{\hat{D}_q(\beta)\hat{\Pi}_q\hat{D}_q(-\beta)}{\Phi_{\text{even}}(t)},
\end{equation}
where $X_1 \equiv \text{Re}(\beta)$ and $X_2 \equiv \text{Im}(\beta)$, and $\hat{\Pi}$ is the parity operator acting on the $q$th harmonic order. Provided that the harmonic spectral amplitudes change slowly with $\alpha$, and setting $\varphi_2(\alpha) = 0$, we find~\cite{rivera-dean_attosecond_2026}
\begin{equation}
	W_q(\beta)
		= \int \dd^2  \alpha \  \abs{c_{\text{BSV}}(\alpha)}^2
				e^{-2\abs{\beta - \chi_q(\alpha)}^2}.
\end{equation}

\begin{figure}
	\centering
	\includegraphics[width=1\columnwidth]{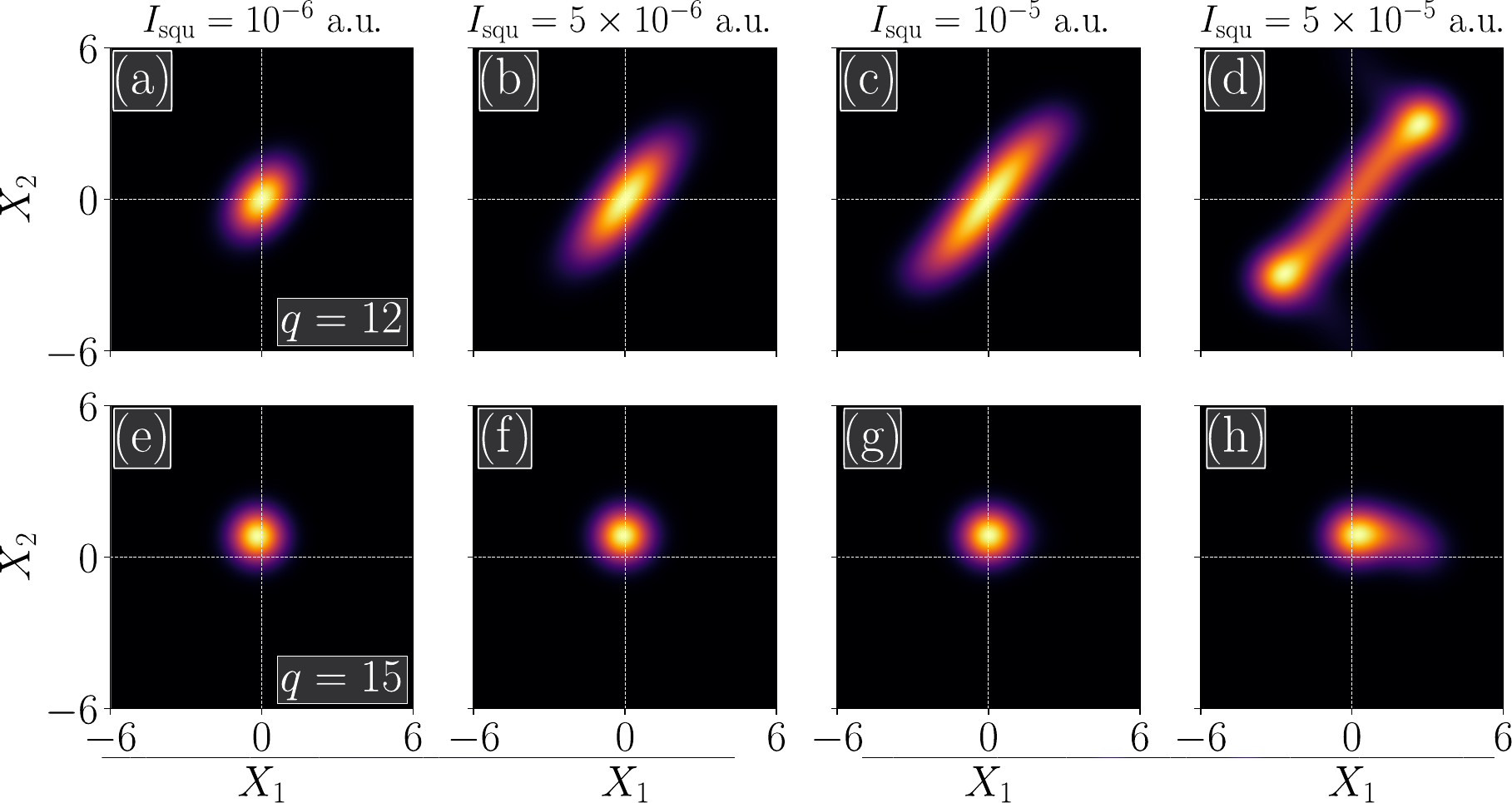}
	\caption{Wigner functions for an even harmonic mode (upper row, $q=12$) and an odd harmonic mode (lower row, $q=15$) for different values of the squeezing intensity, increasing from left to right. The same driving-field configuration and atomic system as in Fig.~\ref{Fig:Bch:fits} have been used.}
	\label{Fig:Wigners}
\end{figure}

Figure~\ref{Fig:Wigners} displays the resulting Wigner functions for an even harmonic mode (upper row) and for an odd harmonic mode (lower row) for different values of the squeezing intensity $I_{\text{squ}} = \kappa^2 \sinh[2](r)$.~For the even harmonic modes, increasing the squeezing strength (from left to right) drives the state from a near-vacuum configuration to a squeezed-like state with reduced uncertainty along a given quadrature~[panels (a)-(c)] that, however, does not decrease as $I_{\text{squ}}$ grows, but instead saturate the vacuum level in agreement with Fig.~\ref{Fig:Eigenvalues:covariance}~(b).~Once the squeezing intensity becomes sufficiently large [panel~(d)], the squeezed-like structure disappears entirely, giving rise to a highly non-Gaussian state characterized by two maxima located approximately symmetrically displaced around the phase-space origin, signaling the breakdown of Gaussianity. 

On the other hand, the odd harmonic modes display a markedly different behavior.~Their Wigner function remains largely unaffected by the squeezed driver at low squeezing intensities [panels (e)-(g)], consistent with the separable nature between even and odd harmonic sectors established in Eq.~\eqref{Eq:separable}.~Then, at high $I_{\text{squ}}$ [panel (h)], develops a pronounced non-Gaussian tail precisely in the regime where the even harmonic modes also lose their Gaussian character.~These observations therefore highlight two distinct regimes:~at low $I_{\text{squ}}$, the generated states remain Gaussian and the even and odd harmonic sectors are effectively decoupled; at high $I_{\text{squ}}$, Gaussianity breaks down simultaneously in both sectors, accompanied by the emergence of correlations between the even and odd harmonic modes.~With the numerical analysis done in this work, and using the regime of applicability derived in the Supplementary Material~\ref{Sec:SM:Gaussian:valid}, we find that $\gamma \sigma \sim 1$ at around $I_{\text{squ}} \simeq 3.6 \times 10^{-5}$ a.u., corresponding to squeezing strengths of $r\simeq 14$, in agreement with the departure from Gaussianity witnessed from the Wigner functions in Fig.~\ref{Fig:Wigners}.~However, despite the non-Gaussian nature of the states in panels (d) and (h), these remain classical in virtue of Hudson's theorem~\cite{hudson_when_1974} due to the absence of negative regions.

It is worth comparing these results with those recently reported in Ref.~\cite{tzur_attosecond-resolved_2025}, where the authors reconstructed Husimi-like distributions of the generated harmonic states using a quantum tomography-like technique~\cite{rivera-dean_attosecond_2026}.~Although the driving field configuration considered there differs slightly from ours, involving a coherent field at frequency $\omega$ and a BSV at frequency $\omega/2$, the reconstructed phase-space distributions exhibit features qualitatively similar to those shown in panels (d) and (h).~More specifically, harmonic modes present in the absence of the BSV field acquire the non-Gaussian tail reported in panel (h), whereas those generated by the squeezed $\omega/2$ component develop a two-peaked structure analogous to that shown in panel (d).~This qualitative agreement suggests that the Gaussianity breakdown predicted here is already accessible with current state-of-the-art BSV sources, hence supporting the feasibility of engineering genuinely entangled Gaussian multimode harmonic states within experimentally realistic driving field configurations.

Nevertheless, it is important to emphasize that the generality of the Gaussian treatment discussed here requires that all harmonics exhibit a spectral amplitude dependence compatible with condition (2) in Sec.~\ref{Sec:Conds:Gauss}. This is a requirement that no longer holds for some of the harmonics obtained using a field configuration where $\omega/2$ is the perturbative BSV and $\omega$ is the strong coherent field, displaying $g^{(2)}>3$, in contrast to the $g^{(2)} = 3$ expected in our case (see Supplementary Material~\ref{Sec:SM:g2}). However, when using the same frequency ratio, we expect this analysis to remain valid under other field-polarization configurations, such as bicircular ones, where the role of the squeezing is to to break the bicircular symmetry of the driving field, enabling symmetry-forbidden harmonics that display $g^{(2)}(0)=3$, while classically-allowed orders display $g^{(2)}(0)=1$~\cite{stammer_fluctuation-induced_2026}, consistent with the same type of even/odd dichotomy found in the Supplementary Material~\ref{Sec:SM:g2}. Finally, the same  $g^{(2)}(0)=3$ and $g^{(2)}(0)=1$ dichotomy also appears to hold in some solid-state materials HHG under similar field configurations, as shown for graphene and MoS$_2$ in Ref.~\cite{stammer_symmetry_2026}.

\section{DISCUSSION}
This work has investigated the conditions under which HHG driven by non-classical fields generates Gaussian quantum states of light, identifying bichromatic driving field configurations composed of a strong coherent field at frequency $\omega$ and a perturbative BSV field at frequency $2\omega$ as a relevant scheme for generating non-trivial multimode Gaussian harmonic states~[Fig.~\ref{Fig:Bch:fits}].~Employing an exact single-quadrature expansion of the BSV driver~[Fig.~\ref{Fig:Fid}], we derived analytical expressions of the covariance matrix associated with arbitrary bipartitions of the generated harmonic modes.~This allowed us to characterize both the collective-mode structure and the single-mode properties of the resulting state~[Fig.~\ref{Fig:Eigenvalues:covariance}], as well as its entanglement structure using tools from Gaussian continuous variable quantum information~[Fig.~\ref{Fig:Ent:DSI}].~We found that the entanglement between a single harmonic mode and the remaining sector is significantly stronger than the pairwise entanglement shared between individual harmonic modes.~Investigating the operational significance of this distinction, we showed that only collective mode bipartitions yield a practically useful quantum advantage in continuous variable teleportation protocols, whereas pairwise harmonic correlations provide at most a marginal improvement over classical strategies~[Fig.~\ref{Fig:fid:teleport}].~Finally, by analyzing the Wigner functions of the generated states, we identified the squeezing regime where the Gaussian description breaks down~[Fig.~\ref{Fig:Wigners}], finding qualitative agreement with recent experimental reconstructions reported in the literature~\cite{tzur_attosecond-resolved_2025}.

From a practical perspective, the entanglement analysis performed in this work requires reconstructing the covariance matrix of the state.~While in our theoretical model $\Gamma_{\mathsf{A}}$ and $\Gamma_{\mathsf{B}}$ are diagonal, environmental effects and experimental imperfections can break this alignment in practice, making full measurement of all covariance matrix elements necessary to reliably certify entanglement in practice.~Thus, simpler evaluation methods are desirable.~Here, we considered the Duan-Simon criterion~\cite{duan_inseparability_2000,simon_peres-horodecki_2000}, which significantly reduces the number of required elements, allowing one to certify entanglement without computing the logarithmic negativity explicitly, at the cost of reduced sensitivity to weakly entangled states.~However, this choice is not unique, and one could further reduce the experimental overhead through convex optimization methods such as semidefinite programming~\cite{BoydCH4}.~These provide necessary conditions for separability by imposing positive semidefinite constraints on the measured moment matrix, an approach that has been similarly leveraged in the discrete variable regime to construct scalable entanglement witnesses~\cite{monteiro_revealing_2015,caspar_local_2022}.~Such techniques offer the possibility of witnessing entanglement without full covariance matrix reconstruction, possibly allowing one to establish a hierarchy of entanglement witnesses based on the required covariance matrix elements~\cite{shchukin_inseparability_2005,navascues_bounding_2007,navascues_convergent_2008}.~This could additionally help identify which mode bipartitions (whether global, pairwise, or an intermediate regime) are most amenable to experimental implementation.

We highlight, however, that covariance matrix reconstruction of the harmonic radiation is, at present, a significant experimental challenge, and remains more of a future perspective than an immediate practical application.~While entanglement witnessing has been carried out in the strong-field context by means of second-order (auto- and cross-) correlation measurements~\cite{theidel_evidence_2024}, these do not, in principle, certify entanglement in our configuration (see Supplementary Material~\ref{Sec:SM:g2}).~One is therefore left needing full or partial covariance matrix reconstructions, which in turn would require engineering local oscillators matched to the specific collective mode of interest, analogous to local oscillator shaping techniques used in the characterization of multimode squeezing from broadband sources~\cite{roslund_wavelength-multiplexed_2014}. An immediate obstacle here is the lack of beam-splitters and other linear optical elements operating at the required frequency regimes, making near-term alternatives based on strong-field techniques, such as using the two-color delay as an effective phase reference in place of a literal local oscillator~\cite{tzur_attosecond-resolved_2025,rivera-dean_attosecond_2026}, a potentially more accessible approach.~Using this idea, standard strong-field pumb-probe techniques using IR and XUV pulses, such as RABBIT~\cite{paul_observation_2001} and attosecond streaking~\cite{hentschel_attosecond_2001,goulielmakis_direct_2004}, have recently been extended to the quantum optical regime~\cite{dubois_quantum_2026,stammer_attosecond_2026} with the aim of characterizing the properties of the driving sources using the generated high-energy photoelectrons.

Finally, while our analysis has mainly focused on Gaussian correlations arising in HHG, we have also identified the conditions under which non-Gaussian states of light can be generated using non-classical Gaussian drivers.~Pure non-Gaussian states are intrinsically non-classical~\cite{hudson_when_1974} and constitute a necessary resource for quantum advantage in applications such as quantum computation~\cite{mari_positive_2012,rahimi-keshari_sufficient_2016}.~In the context of HHG, and when focusing on single modes, such non-Gaussian features manifest as mixed states with extreme super-Poissonian statistics~\cite{petrovic_generation_2026}, arising from the combination of the broad BSV phase-space distribution and the high nonlinearity of the HHG process~[Fig.~\ref{Fig:BSV:fits}].~These results further position HHG, and more broadly strong-field physics processes, driven by non-classical light as a plausible platform for the generation of multimode non-Gaussian entangled states of light~\cite{rivera-dean_erasing_2026} in the high-intensity domain, spaning frequencies from the near infrared to the extreme ultraviolet regimes, and with entanglement structure and non-classical properties remaining as open questions for further investigation.

\section*{ACKNOWLEDGMENTS}
J.~R.-D.~gratefully acknowledges Margarita Khokhlova and Emilio Pisanty for their kind hospitality at King's College London and many fruitful discussions, as well as Lidija Petrovic for insightful discussions.

J.~R.-D.~acknowledges funding from UK Engineering and Physical Sciences Research Council (EPSRC) Funding, Grant UKRI2300 - Attosecond Photoelectron Imaging with Quantum Light (APIQuL).

ICFO-QOT group acknowledges support from:
MCIN/AEI (PGC2018-0910.13039/501100011033, CEX2019-000910-S/10.13039/501100011033, Plan National STAMEENA PID2022-139099NB, project funded MCIN and by the ``European Union NextGenerationEU/PRTR'' (PRTR-C17.I1), FPI); Ministry for Digital Transformation and of Civil Service of the Spanish Government through the QUANTUM ENIA project call - Quantum Spain project, and by the European Union through the Recovery, Transformation and Resilience Plan - NextGenerationEU within the framework of the Digital Spain 2026 Agenda; CEX2024-001490-S [MICIU/AEI/10.13039/501100011033]; Fundació Cellex;
Fundació Mir-Puig; Generalitat de Catalunya (European Social Fund FEDER and CERCA program; Barcelona Supercomputing Center MareNostrum (FI-2023-3-0024);
Funded by the European Union (HORIZON-CL4-2022-QUANTUM-02-SGA, PASQuanS2.1, 101113690, EU Horizon 2020 FET-OPEN OPTOlogic, Grant No 899794, QU-ATTO, 101168628), EU Horizon Europe Program (No 101080086 NeQSTGrant Agreement 101080086 — NeQST).

O.~C.~and M.~F.~C.~acknowledge support by the Quantum Science and Technology-National Science and Technology Major Project (Grant No.~2025ZD0301000), the Guangdong Provincial Quantum Science Strategic Initiative (Grant No.~GDZX2504001), the National Key Research and Development Program of China (Grant No.~2023YFA1407100), the Guangdong Province Science and Technology Major Project (Future functional materials under extreme conditions - 2021B0301030005) and the National Natural Science Foundation of China (Grant No.~12574092).

\bibliography{References.bib}{}

@article{hentschel_attosecond_2001,
	title = {Attosecond metrology},
	volume = {414},
	copyright = {2001 Macmillan Magazines Ltd.},
	issn = {1476-4687},
	url = {https://www.nature.com/articles/35107000},
	doi = {10.1038/35107000},
	number = {6863},
	urldate = {2026-08-22},
	journal = {Nature},
	author = {Hentschel, M. and Kienberger, R. and Spielmann, Ch and Reider, G. A. and Milosevic, N. and Brabec, T. and Corkum, P. and Heinzmann, U. and Drescher, M. and Krausz, F.},
	month = nov,
	year = {2001},
	pages = {509--513},
}

@misc{stammer_attosecond_2026,
	title = {Attosecond metrology of bright quantum light},
	url = {http://arxiv.org/abs/2607.06395},
	doi = {10.48550/arXiv.2607.06395},
	urldate = {2026-08-22},
	publisher = {arXiv},
	author = {Stammer, P. and Rivera-Dean, J. and Pisanty, E. and Lewenstein, M.},
	month = jul,
	year = {2026},
	note = {arXiv:2607.06395 [quant-ph]},
}

@misc{dubois_quantum_2026,
	title = {Quantum optical photoelectron interferometry},
	url = {http://arxiv.org/abs/2606.13447},
	doi = {10.48550/arXiv.2606.13447},
	urldate = {2026-08-22},
	publisher = {arXiv},
	author = {Dubois, Jonathan and Cotte, Viviane and Taïeb, Richard and Lévêque, Camille and Caillat, Jérémie and Dave, Pranshu and Salières, Pascal and Bresteau, David and Bourassin-Bouchet, Charles and L'Huillier, Anne and Busto, David},
	month = jun,
	year = {2026},
	note = {arXiv:2606.13447 [quant-ph]},
}

@article{goulielmakis_direct_2004,
	title = {Direct {Measurement} of {Light} {Waves}},
	volume = {305},
	url = {https://www.science.org/doi/abs/10.1126/science.1100866},
	doi = {10.1126/science.1100866},
	number = {5688},
	urldate = {2023-11-30},
	journal = {Science},
	publisher = {American Association for the Advancement of Science},
	author = {Goulielmakis, E. and Uiberacker, M. and Kienberger, R. and Baltuska, A. and Yakovlev, V. and Scrinzi, A. and Westerwalbesloh, Th. and Kleineberg, U. and Heinzmann, U. and Drescher, M. and Krausz, F.},
	month = aug,
	year = {2004},
	pages = {1267--1269},
}

@misc{stammer_symmetry_2026,
	title = {Symmetry breaking by quantum light in solid-state high-harmonic generation},
	url = {http://arxiv.org/abs/2605.28236},
	doi = {10.48550/arXiv.2605.28236},
	urldate = {2026-08-20},
	publisher = {arXiv},
	author = {Stammer, P. and Rivera-Dean, J. and Kim, D. and Chacón, A. and Gao, W. and Granados, C.},
	month = may,
	year = {2026},
	note = {arXiv:2605.28236 [physics.optics]},
}

@misc{wang_generation_2026,
	title = {Generation of bright quantum high-order harmonic driven by combined coherent and bright squeezed vacuum light},
	url = {http://arxiv.org/abs/2607.20941},
	doi = {10.48550/arXiv.2607.20941},
	urldate = {2026-08-19},
	publisher = {arXiv},
	author = {Wang, Wentao and Sun, Yaoshun and Wang, Liyuan and Hu, Lingrui and Ding, Dajun and Tang, Xiangyu and Li, Mingxuan and Yuan, Jianmin and Luo, Sizuo},
	month = jul,
	year = {2026},
	note = {arXiv:2607.20941 [quant-ph]},
}

@article{rivera-dean_propagation_2026,
	title = {Propagation effects in high harmonic generation media driven by bright squeezed vacuum light},
	volume = {17},
	copyright = {2026 The Author(s)},
	issn = {2041-1723},
	url = {https://www.nature.com/articles/s41467-026-75173-8},
	doi = {10.1038/s41467-026-75173-8},
	number = {1},
	urldate = {2026-08-19},
	journal = {Nature Communications},
	author = {Rivera-Dean, Javier and Kanti, Deeksha and Stammer, Philipp and Carlström, Stefanos and Tsatrafyllis, Nikolaos and Ivanov, Misha Yu and Lewenstein, Maciej and Tzallas, Paraskevas},
	month = jul,
	year = {2026},
	pages = {8457},
}

@article{sundaram_high-order_1990,
	title = {High-order harmonic generation: {Simplified} model and relevance of single-atom theories to experiment},
	volume = {41},
	shorttitle = {High-order harmonic generation},
	url = {https://link.aps.org/doi/10.1103/PhysRevA.41.6571},
	doi = {10.1103/PhysRevA.41.6571},
	number = {11},
	urldate = {2023-01-25},
	journal = {Physical Review A},
	publisher = {American Physical Society},
	author = {Sundaram, Bala and Milonni, Peter W.},
	month = jun,
	year = {1990},
	pages = {6571--6573},
}

@misc{rivera-dean_erasing_2026,
	title = {Erasing photons from bright squeezed vacuum light via above-threshold ionization},
	copyright = {arXiv.org perpetual, non-exclusive license},
	url = {https://arxiv.org/abs/2605.31160},
	doi = {10.48550/ARXIV.2605.31160},
	urldate = {2026-06-03},
	publisher = {arXiv},
	author = {Rivera-Dean, J. and Rook, T. and Singh, G. and Stammer, P. and Khokhlova, M. and Pisanty, E. and Faria, C. Figueira de Morisson},
	year = {2026},
	note = {arXiv:2605.31160 [quant-ph]},
}

@article{navascues_bounding_2007,
	title = {Bounding the {Set} of {Quantum} {Correlations}},
	volume = {98},
	url = {https://link.aps.org/doi/10.1103/PhysRevLett.98.010401},
	doi = {10.1103/PhysRevLett.98.010401},
	number = {1},
	urldate = {2022-06-11},
	journal = {Physical Review Letters},
	publisher = {American Physical Society},
	author = {Navascués, Miguel and Pironio, Stefano and Acín, Antonio},
	month = jan,
	year = {2007},
	pages = {010401},
}

@article{navascues_convergent_2008,
	title = {A convergent hierarchy of semidefinite programs characterizing the set of quantum correlations},
	volume = {10},
	issn = {1367-2630},
	url = {https://dx.doi.org/10.1088/1367-2630/10/7/073013},
	doi = {10.1088/1367-2630/10/7/073013},
	number = {7},
	urldate = {2023-12-04},
	journal = {New Journal of Physics},
	author = {Navascués, Miguel and Pironio, Stefano and Acín, Antonio},
	month = jul,
	year = {2008},
	pages = {073013},
}

@misc{RBSFA,
	author = {Pisanty, Emilio},
	title = {{RB-SFA: High Harmonic Generation in the Strong Field Approximation via Mathematica}},
	howpublished = {Github: \url{https://github.com/episanty/RB-SFA}},
	doi = {10.5281/zenodo.592519},
	year = {2020}
}

@article{lange_hierarchy_2025,
	title = {Hierarchy of approximations for describing quantum light from high-harmonic generation: {A} {Fermi}-{Hubbard}-model study},
	volume = {111},
	issn = {2469-9926, 2469-9934},
	shorttitle = {Hierarchy of approximations for describing quantum light from high-harmonic generation},
	url = {https://link.aps.org/doi/10.1103/PhysRevA.111.013113},
	doi = {10.1103/PhysRevA.111.013113},
	number = {1},
	urldate = {2025-03-05},
	journal = {Physical Review A},
	author = {Lange, Christian Saugbjerg and Madsen, Lars Bojer},
	month = jan,
	year = {2025},
	pages = {013113},
}

@article{rivera-dean_squeezed_2024,
	title = {Squeezed states of light after high-order harmonic generation in excited atomic systems},
	volume = {110},
	url = {https://link.aps.org/doi/10.1103/PhysRevA.110.063118},
	doi = {10.1103/PhysRevA.110.063118},
	number = {6},
	urldate = {2025-01-06},
	journal = {Physical Review A},
	publisher = {American Physical Society},
	author = {Rivera-Dean, J. and Crispin, H. B. and Stammer, P. and Lamprou, Th. and Pisanty, E. and Krüger, M. and Tzallas, P. and Lewenstein, M. and Ciappina, M. F.},
	month = dec,
	year = {2024},
	pages = {063118},
}

@inbook{ScullyBookCh2,
	author = {Scully, Marlan O. and Zubairy, M. Suhail},
	publisher = {Cambridge University Press, Cambridge, UK},
	isbn = {0-521-43458-0},
	title = {{Coherent and squeezed states of the radiation fieldy}},
	booktitle = {{Quantum Optics}},
	chapter = {2},
	pages = {1-45},
	year = {2001},
}

@inbook{GerryBookCh7,
	author = {C. Gerry and P. Knight},
	publisher = {Cambridge University Press, Cambridge, UK},
	isbn = {978-0-521-52735-4},
	title = {{Nonclassical light}},
	booktitle = {{Introductory Quantum Optics}},
	chapter = {7},
	pages = {150-194},
	year = {2005},
}

@incollection{jaynes_central_2003,
	location = {Cambridge},
	title = {The central, Gaussian or normal distribution},
	isbn = {978-0-521-59271-0},
	url = {https://www.cambridge.org/core/books/probability-theory/central-gaussian-or-normal-distribution/A4DFF7BDCC283F050BF4D88421E30545},
	doi = {10.1017/CBO9780511790423.009},
	pages = {198--242},
	booktitle = {Probability Theory: The Logic of Science},
	publisher = {Cambridge University Press},
	author = {Jaynes, E. T. and Bretthorst, G. Larry},
	urldate = {2026-05-18},
	year = {2003},
}

@incollection{serafini_entanglement_2023,
	edition = {2},
	title = {Entanglement of Continuous Variable Systems},
	booktitle = {Quantum Continuous Variables},
	publisher = {{CRC} Press},
	author = {Serafini, Alessio},
	year = {2023},
}

@incollection{serafini_phase_2023,
	edition = {2},
	title = {Phase Space Methods},
	booktitle = {Quantum Continuous Variables},
	publisher = {{CRC} Press},
	author = {Serafini, Alessio},
	year = {2023},
}

@incollection{serafini_gaussian_2023,
	edition = {2},
	title = {Gaussian Operations},
	booktitle = {Quantum Continuous Variables},
	publisher = {{CRC} Press},
	author = {Serafini, Alessio},
	year = {2023},
}

@misc{stammer_weak_2025,
	title = {Weak measurement in strong laser field physics},
	url = {http://arxiv.org/abs/2508.09048},
	doi = {10.48550/arXiv.2508.09048},
	urldate = {2025-09-05},
	publisher = {arXiv},
	author = {Stammer, Philipp and Rivera-Dean, Javier and Ciappina, Marcelo F. and Lewenstein, Maciej},
	month = aug,
	year = {2025},
	note = {arXiv:2508.09048 [quant-ph]},
}

@article{tzur_generation_2024,
	title = {Generation of squeezed high-order harmonics},
	volume = {6},
	url = {https://link.aps.org/doi/10.1103/PhysRevResearch.6.033079},
	doi = {10.1103/PhysRevResearch.6.033079},
	pages = {033079},
	number = {3},
	journal = {Physical Review Research},
	shortjournal = {Phys. Rev. Res.},
	publisher = {American Physical Society},
	author = {Tzur, Matan Even and Birk, Michael and Gorlach, Alexey and Kaminer, Ido and Krüger, Michael and Cohen, Oren},
	urldate = {2024-07-23},
	year = {2024},
}

@article{krause_high-order_1992,
	title = {High-order harmonic generation from atoms and ions in the high intensity regime},
	volume = {68},
	url = {https://link.aps.org/doi/10.1103/PhysRevLett.68.3535},
	doi = {10.1103/PhysRevLett.68.3535},
	number = {24},
	urldate = {2022-07-20},
	journal = {Physical Review Letters},
	author = {Krause, Jeffrey L. and Schafer, Kenneth J. and Kulander, Kenneth C.},
	month = jun,
	year = {1992},
	pages = {3535--3538},
}

@article{corkum_plasma_1993,
	title = {Plasma perspective on strong field multiphoton ionization},
	volume = {71},
	url = {https://link.aps.org/doi/10.1103/PhysRevLett.71.1994},
	doi = {10.1103/PhysRevLett.71.1994},
	number = {13},
	urldate = {2022-01-16},
	journal = {Physical Review Letters},
	author = {Corkum, P. B.},
	month = sep,
	year = {1993},
	pages = {1994--1997},
}

@article{heidmann_observation_1987,
	title = {Observation of Quantum Noise Reduction on Twin Laser Beams},
	volume = {59},
	url = {https://link.aps.org/doi/10.1103/PhysRevLett.59.2555},
	doi = {10.1103/PhysRevLett.59.2555},
	pages = {2555--2557},
	number = {22},
	journal = {Physical Review Letters},
	publisher = {American Physical Society},
	author = {Heidmann, A. and Horowicz, R. J. and Reynaud, S. and Giacobino, E. and Fabre, C. and Camy, G.},
	urldate = {2026-05-18},
	year = {1987},
}

@article{ou_realization_1992,
	title = {Realization of the Einstein-Podolsky-Rosen paradox for continuous variables},
	volume = {68},
	url = {https://link.aps.org/doi/10.1103/PhysRevLett.68.3663},
	doi = {10.1103/PhysRevLett.68.3663},
	pages = {3663--3666},
	number = {25},
	journal = {Physical Review Letters},
	publisher = {American Physical Society},
	author = {Ou, Z. Y. and Pereira, S. F. and Kimble, H. J. and Peng, K. C.},
	urldate = {2026-05-18},
	year = {1992},
}

@misc{lyu_attosecond_2026,
	title = {Attosecond quantum spectroscopy with entangled photon pairs},
	rights = {{arXiv}.org perpetual, non-exclusive license},
	url = {https://arxiv.org/abs/2604.06707},
	doi = {10.48550/ARXIV.2604.06707},
	number = {{arXiv}:2604.06707},
	publisher = {{arXiv}},
	author = {Lyu, Zijian and Sun, Fengxiao and Yi, Sili and Li, Jingze and Liu, Haodong and He, Qiongyi and Gong, Qihuang and Ivanov, Misha and Liu, Yunquan},
	urldate = {2026-05-18},
	year = {2026},
}

@article{kulander_theory_1990,
	title = {Theory of {High}-{Order} {Processes} in {Atoms} in {Intense} {Laser} {Fields}: {Introduction}},
	volume = {7},
	issn = {1520-8540},
	shorttitle = {Theory of {High}-{Order} {Processes} in {Atoms} in {Intense} {Laser} {Fields}},
	url = {https://opg.optica.org/josab/abstract.cfm?uri=josab-7-4-403},
	doi = {10.1364/JOSAB.7.000403},
	number = {4},
	urldate = {2025-10-27},
	journal = {JOSA B},
	author = {Kulander, Kenneth and L’Huillier, Anne},
	month = apr,
	year = {1990},
	pages = {403--402},
}

@article{lhuillier_calculations_1992,
	title = {Calculations of high-order harmonic-generation processes in xenon at 1064 nm},
	volume = {46},
	copyright = {http://link.aps.org/licenses/aps-default-license},
	issn = {1050-2947, 1094-1622},
	url = {https://link.aps.org/doi/10.1103/PhysRevA.46.2778},
	doi = {10.1103/PhysRevA.46.2778},
	number = {5},
	urldate = {2025-10-27},
	journal = {Physical Review A},
	author = {L’Huillier, Anne and Balcou, Philippe and Candel, Sebastien and Schafer, Kenneth J. and Kulander, Kenneth C.},
	month = sep,
	year = {1992},
	pages = {2778--2790},
}

@article{weissenbilder_how_2022,
	title = {How to optimize high-order harmonic generation in gases},
	volume = {4},
	copyright = {2022 Springer Nature Limited},
	issn = {2522-5820},
	url = {https://www.nature.com/articles/s42254-022-00522-7},
	doi = {10.1038/s42254-022-00522-7},
	number = {11},
	urldate = {2024-10-01},
	journal = {Nature Reviews Physics},
	author = {Weissenbilder, R. and Carlström, S. and Rego, L. and Guo, C. and Heyl, C. M. and Smorenburg, P. and Constant, E. and Arnold, C. L. and L’Huillier, A.},
	month = nov,
	year = {2022},
	pages = {713--722},
}

@article{eberly_spectrum_1992,
	title = {Spectrum of light scattered coherently or incoherently by a collection of atoms},
	volume = {45},
	url = {https://link.aps.org/doi/10.1103/PhysRevA.45.4706},
	doi = {10.1103/PhysRevA.45.4706},
	pages = {4706--4712},
	number = {7},
	journal = {Physical Review A},
	shortjournal = {Phys. Rev. A},
	publisher = {American Physical Society},
	author = {Eberly, J. H. and Fedorov, M. V.},
	urldate = {2023-05-10},
	year = {1992},
}

@article{rivera-dean_quantum-optical_2024,
	title = {Quantum-optical analysis of high-order harmonic generation in {H}$_2^+$ molecules},
	volume = {109},
	url = {https://link.aps.org/doi/10.1103/PhysRevA.109.033706},
	doi = {10.1103/PhysRevA.109.033706},
	pages = {033706},
	number = {3},
	journal = {Physical Review A},
	shortjournal = {Phys. Rev. A},
	publisher = {American Physical Society},
	author = {Rivera-Dean, J. and Stammer, P. and Maxwell, A. S. and Lamprou, Th. and Pisanty, E. and Tzallas, P. and Lewenstein, M. and Ciappina, M. F.},
	urldate = {2024-03-19},
	year = {2024},
}

@misc{stammer_theory_2025,
	title = {Theory of quantum optics and optical coherence in high harmonic generation},
	url = {http://arxiv.org/abs/2504.13287},
	doi = {10.48550/arXiv.2504.13287},
	number = {{arXiv}:2504.13287},
	publisher = {{arXiv}},
	author = {Stammer, Philipp and Rivera-Dean, Javier and Lewenstein, Maciej},
	urldate = {2025-06-09},
	year = {2025},
	eprinttype = {arxiv},
	eprint = {2504.13287 [quant-ph]},
}

@article{stammer_theory_2022,
	title = {Theory of entanglement and measurement in high-order harmonic generation},
	volume = {106},
	url = {https://link.aps.org/doi/10.1103/PhysRevA.106.L050402},
	doi = {10.1103/PhysRevA.106.L050402},
	pages = {L050402},
	number = {5},
	journal = {Physical Review A},
	shortjournal = {Phys. Rev. A},
	publisher = {American Physical Society},
	author = {Stammer, Philipp},
	urldate = {2023-02-09},
	year = {2022},
}

@article{lai_characteristic_1989,
	title = {Characteristic functions and quantum measurements of optical observables},
	volume = {1},
	issn = {0954-8998},
	url = {https://doi.org/10.1088/0954-8998/1/2/003},
	doi = {10.1088/0954-8998/1/2/003},
	pages = {99},
	number = {2},
	journal = {Quantum Optics: Journal of the European Optical Society Part B},
	shortjournal = {Quantum Opt.},
	author = {Lai, Y. and Haus, H. A.},
	urldate = {2026-05-18},
	year = {1989},
}

@incollection{vogel_bosonic_2006,
	title = {Bosonic systems in phase space},
	rights = {Copyright © 2006 Wiley-{VCH} Verlag {GmbH} \& Co. {KGaA}},
	isbn = {978-3-527-60852-2},
	url = {https://onlinelibrary.wiley.com/doi/abs/10.1002/3527608524.ch4},
	doi = {10.1002/3527608524.ch4},
	author = {Vogel, Werner and Welsch, Dirk‐Gunnar},
	pages = {113--134},
	booktitle = {Quantum Optics},
	publisher = {John Wiley \& Sons, Ltd},
	urldate = {2026-05-18},
	year = {2006},
}

@article{dahlstrom_quantum_2011,
	title = {Quantum mechanical approach to probing the birth of attosecond pulses using a two-colour field},
	volume = {44},
	issn = {0953-4075, 1361-6455},
	url = {https://iopscience.iop.org/article/10.1088/0953-4075/44/9/095602},
	doi = {10.1088/0953-4075/44/9/095602},
	number = {9},
	urldate = {2025-02-24},
	journal = {Journal of Physics B: Atomic, Molecular and Optical Physics},
	author = {Dahlström, J M and L'Huillier, A and Mauritsson, J},
	month = may,
	year = {2011},
	pages = {095602},
}

@article{pedatzur_attosecond_2015,
	title = {Attosecond tunnelling interferometry},
	volume = {11},
	rights = {2015 Springer Nature Limited},
	issn = {1745-2481},
	url = {https://www.nature.com/articles/nphys3436},
	doi = {10.1038/nphys3436},
	pages = {815--819},
	number = {10},
	journal = {Nature Physics},
	publisher = {Nature Publishing Group},
	author = {Pedatzur, O. and Orenstein, G. and Serbinenko, V. and Soifer, H. and Bruner, B. D. and Uzan, A. J. and Brambila, D. S. and Harvey, A. G. and Torlina, L. and Morales, F. and Smirnova, O. and Dudovich, N.},
	urldate = {2024-04-05},
	year = {2015},
	langid = {english},
}

@article{cai_multimode_2017,
	title = {Multimode entanglement in reconfigurable graph states using optical frequency combs},
	volume = {8},
	copyright = {2017 The Author(s)},
	issn = {2041-1723},
	url = {https://www.nature.com/articles/ncomms15645},
	doi = {10.1038/ncomms15645},
	number = {1},
	urldate = {2026-05-18},
	journal = {Nature Communications},
	publisher = {Nature Publishing Group},
	author = {Cai, Y. and Roslund, J. and Ferrini, G. and Arzani, F. and Xu, X. and Fabre, C. and Treps, N.},
	month = jun,
	year = {2017},
	pages = {15645},
}

@article{chen_experimental_2014,
	title = {Experimental {Realization} of {Multipartite} {Entanglement} of 60 {Modes} of a {Quantum} {Optical} {Frequency} {Comb}},
	volume = {112},
	url = {https://link.aps.org/doi/10.1103/PhysRevLett.112.120505},
	doi = {10.1103/PhysRevLett.112.120505},
	number = {12},
	urldate = {2026-05-18},
	journal = {Physical Review Letters},
	publisher = {American Physical Society},
	author = {Chen, Moran and Menicucci, Nicolas C. and Pfister, Olivier},
	month = mar,
	year = {2014},
	pages = {120505},
}

@article{van_loock_building_2007,
	title = {Building {Gaussian} cluster states by linear optics},
	volume = {76},
	url = {https://link.aps.org/doi/10.1103/PhysRevA.76.032321},
	doi = {10.1103/PhysRevA.76.032321},
	number = {3},
	urldate = {2026-05-18},
	journal = {Physical Review A},
	publisher = {American Physical Society},
	author = {van Loock, Peter and Weedbrook, Christian and Gu, Mile},
	month = sep,
	year = {2007},
	pages = {032321},
}

@article{gerke_full_2015,
	title = {Full {Multipartite} {Entanglement} of {Frequency}-{Comb} {Gaussian} {States}},
	volume = {114},
	url = {https://link.aps.org/doi/10.1103/PhysRevLett.114.050501},
	doi = {10.1103/PhysRevLett.114.050501},
	number = {5},
	urldate = {2026-05-18},
	journal = {Physical Review Letters},
	publisher = {American Physical Society},
	author = {Gerke, S. and Sperling, J. and Vogel, W. and Cai, Y. and Roslund, J. and Treps, N. and Fabre, C.},
	month = feb,
	year = {2015},
	pages = {050501},
}

@article{horodecki_separability_1996,
	title = {Separability of mixed states: necessary and sufficient conditions},
	volume = {223},
	issn = {0375-9601},
	shorttitle = {Separability of mixed states},
	url = {https://www.sciencedirect.com/science/article/pii/S0375960196007062},
	doi = {10.1016/S0375-9601(96)00706-2},
	number = {1},
	urldate = {2022-01-20},
	journal = {Physics Letters A},
	author = {Horodecki, Michał and Horodecki, Paweł and Horodecki, Ryszard},
	month = nov,
	year = {1996},
	pages = {1--8},
}

@article{peres_separability_1996,
	title = {Separability {Criterion} for {Density} {Matrices}},
	volume = {77},
	url = {https://link.aps.org/doi/10.1103/PhysRevLett.77.1413},
	doi = {10.1103/PhysRevLett.77.1413},
	number = {8},
	urldate = {2022-01-20},
	journal = {Physical Review Letters},
	publisher = {American Physical Society},
	author = {Peres, Asher},
	month = aug,
	year = {1996},
	pages = {1413--1415},
}

@article{medeiros_de_araujo_full_2014,
	title = {Full characterization of a highly multimode entangled state embedded in an optical frequency comb using pulse shaping},
	volume = {89},
	url = {https://link.aps.org/doi/10.1103/PhysRevA.89.053828},
	doi = {10.1103/PhysRevA.89.053828},
	number = {5},
	urldate = {2026-05-18},
	journal = {Physical Review A},
	publisher = {American Physical Society},
	author = {Medeiros de Araújo, R. and Roslund, J. and Cai, Y. and Ferrini, G. and Fabre, C. and Treps, N.},
	month = may,
	year = {2014},
	pages = {053828},
}

@article{roslund_wavelength-multiplexed_2014,
	title = {Wavelength-multiplexed quantum networks with ultrafast frequency combs},
	volume = {8},
	copyright = {2013 Springer Nature Limited},
	issn = {1749-4893},
	url = {https://www.nature.com/articles/nphoton.2013.340},
	doi = {10.1038/nphoton.2013.340},
	number = {2},
	urldate = {2026-05-18},
	journal = {Nature Photonics},
	publisher = {Nature Publishing Group},
	author = {Roslund, Jonathan and de Araújo, Renné Medeiros and Jiang, Shifeng and Fabre, Claude and Treps, Nicolas},
	month = feb,
	year = {2014},
	pages = {109--112},
}

@article{braunstein_quantum_2005,
	title = {Quantum information with continuous variables},
	volume = {77},
	url = {https://link.aps.org/doi/10.1103/RevModPhys.77.513},
	doi = {10.1103/RevModPhys.77.513},
	number = {2},
	urldate = {2022-04-22},
	journal = {Reviews of Modern Physics},
	publisher = {American Physical Society},
	author = {Braunstein, Samuel L. and van Loock, Peter},
	month = jun,
	year = {2005},
	pages = {513--577},
}

@inbook{NielsenBookCh1,
	booktitle = {Quantum {Computation} and {Quantum} {Information}: 10th {Anniversary} {Edition}},
	isbn = {978-1-139-49548-6},
	title = {Introduction and overview},
	publisher = {Cambridge University Press, Cambridge, UK},
	author = {Nielsen, Michael A. and Chuang, Isaac L.},
	month = dec,
	year = {2010},
	chapter = {1},
	pages = {1-59},
}

@article{mari_optimal_2008,
	title = {Optimal fidelity of teleportation of coherent states and entanglement},
	volume = {78},
	url = {https://link.aps.org/doi/10.1103/PhysRevA.78.062340},
	doi = {10.1103/PhysRevA.78.062340},
	number = {6},
	urldate = {2026-05-18},
	journal = {Physical Review A},
	publisher = {American Physical Society},
	author = {Mari, A. and Vitali, D.},
	month = dec,
	year = {2008},
	pages = {062340},
}

@article{braunstein_teleportation_1998,
	title = {Teleportation of {Continuous} {Quantum} {Variables}},
	volume = {80},
	url = {https://link.aps.org/doi/10.1103/PhysRevLett.80.869},
	doi = {10.1103/PhysRevLett.80.869},
	number = {4},
	urldate = {2023-09-17},
	journal = {Physical Review Letters},
	publisher = {American Physical Society},
	author = {Braunstein, Samuel L. and Kimble, H. J.},
	month = jan,
	year = {1998},
	pages = {869--872},
}

@article{adesso_equivalence_2005,
	title = {Equivalence between {Entanglement} and the {Optimal} {Fidelity} of {Continuous} {Variable} {Teleportation}},
	volume = {95},
	url = {https://link.aps.org/doi/10.1103/PhysRevLett.95.150503},
	doi = {10.1103/PhysRevLett.95.150503},
	number = {15},
	urldate = {2026-05-18},
	journal = {Physical Review Letters},
	publisher = {American Physical Society},
	author = {Adesso, Gerardo and Illuminati, Fabrizio},
	month = oct,
	year = {2005},
	pages = {150503},
}

@misc{tzur_attosecond-resolved_2025,
	title = {Attosecond-resolved quantum fluctuations of light and matter},
	copyright = {arXiv.org perpetual, non-exclusive license},
	url = {https://arxiv.org/abs/2511.18362},
	doi = {10.48550/ARXIV.2511.18362},
	urldate = {2026-05-18},
	number = {{arXiv}:2511.18362},
	publisher = {{arXiv}},
	author = {Tzur, Matan Even and Mor, Chen and Yaffe, Noa and Birk, Michael and Rasputnyi, Andrei and Kneller, Omer and Nisim, Ido and Kaminer, Ido and Chekhova, Maria and Krueger, Michael and Ivanov, Misha and Dudovich, Nirit and Cohen, Oren},
	year = {2025},
}

@article{simon_peres-horodecki_2000,
	title = {Peres-{Horodecki} {Separability} {Criterion} for {Continuous} {Variable} {Systems}},
	volume = {84},
	url = {https://link.aps.org/doi/10.1103/PhysRevLett.84.2726},
	doi = {10.1103/PhysRevLett.84.2726},
	number = {12},
	urldate = {2024-07-02},
	journal = {Physical Review Letters},
	publisher = {American Physical Society},
	author = {Simon, R.},
	month = mar,
	year = {2000},
	pages = {2726--2729},
}

@misc{petrovic_generation_2026,
	title = {Generation of circular polarized high-order harmonics from single color quantum light},
	url = {http://arxiv.org/abs/2601.01611},
	doi = {10.48550/arXiv.2601.01611},
	urldate = {2026-01-19},
	publisher = {arXiv},
	author = {Petrovic, Lidija and Stammer, Philipp and Lewenstein, Maciej and Rivera-Dean, Javier},
	month = jan,
	year = {2026},
	note = {arXiv:2601.01611 [quant-ph]},
}

@inbook{BoydCH4,
	booktitle = {{Convex Optimization}},
	isbn = {0 521 83378 7},
	title = {Convex optimization problems},
	publisher = {Cambridge University Press, Cambridge, UK},
	author = {Boyd, Stephen and Lieven, Vandenberghe},
	month = dec,
	year = {2004},
	chapter = {4},
	pages = {127-214},
}

@article{monteiro_revealing_2015,
	title = {Revealing {Genuine} {Optical}-{Path} {Entanglement}},
	volume = {114},
	url = {https://link.aps.org/doi/10.1103/PhysRevLett.114.170504},
	doi = {10.1103/PhysRevLett.114.170504},
	number = {17},
	urldate = {2026-05-18},
	journal = {Physical Review Letters},
	publisher = {American Physical Society},
	author = {Monteiro, F. and Vivoli, V. Caprara and Guerreiro, T. and Martin, A. and Bancal, J.-D. and Zbinden, H. and Thew, R. T. and Sangouard, N.},
	month = may,
	year = {2015},
	pages = {170504},
}

@article{shchukin_inseparability_2005,
	title = {Inseparability {Criteria} for {Continuous} {Bipartite} {Quantum} {States}},
	volume = {95},
	copyright = {http://link.aps.org/licenses/aps-default-license},
	issn = {0031-9007, 1079-7114},
	url = {https://link.aps.org/doi/10.1103/PhysRevLett.95.230502},
	doi = {10.1103/PhysRevLett.95.230502},
	number = {23},
	urldate = {2026-05-18},
	journal = {Physical Review Letters},
	author = {Shchukin, E. and Vogel, W.},
	month = nov,
	year = {2005},
	pages = {230502},
}

@article{duan_inseparability_2000,
	title = {Inseparability {Criterion} for {Continuous} {Variable} {Systems}},
	volume = {84},
	url = {https://link.aps.org/doi/10.1103/PhysRevLett.84.2722},
	doi = {10.1103/PhysRevLett.84.2722},
	number = {12},
	urldate = {2024-07-02},
	journal = {Physical Review Letters},
	publisher = {American Physical Society},
	author = {Duan, Lu-Ming and Giedke, G. and Cirac, J. I. and Zoller, P.},
	month = mar,
	year = {2000},
	pages = {2722--2725},
}

@article{caspar_local_2022,
	title = {Local and scalable detection of genuine multipartite single-photon path entanglement},
	volume = {6},
	url = {https://quantum-journal.org/papers/q-2022-03-22-671/},
	doi = {10.22331/q-2022-03-22-671},
	urldate = {2023-11-25},
	journal = {Quantum},
	publisher = {Verein zur Förderung des Open Access Publizierens in den Quantenwissenschaften},
	author = {Caspar, Patrik and Oudot, Enky and Sekatski, Pavel and Maring, Nicolas and Martin, Anthony and Sangouard, Nicolas and Zbinden, Hugo and Thew, Rob},
	month = mar,
	year = {2022},
	pages = {671},
}

@article{hudson_when_1974,
	title = {When is the wigner quasi-probability density non-negative?},
	volume = {6},
	issn = {0034-4877},
	url = {https://www.sciencedirect.com/science/article/pii/003448777490007X},
	doi = {10.1016/0034-4877(74)90007-X},
	number = {2},
	urldate = {2022-04-20},
	journal = {Reports on Mathematical Physics},
	author = {Hudson, R. L.},
	month = oct,
	year = {1974},
	pages = {249--252},
}

@article{royer_wigner_1977,
	title = {Wigner function as the expectation value of a parity operator},
	volume = {15},
	url = {https://link.aps.org/doi/10.1103/PhysRevA.15.449},
	doi = {10.1103/PhysRevA.15.449},
	number = {2},
	urldate = {2022-06-18},
	journal = {Physical Review A},
	publisher = {American Physical Society},
	author = {Royer, Antoine},
	month = feb,
	year = {1977},
	pages = {449--450},
}

@article{rahimi-keshari_sufficient_2016,
	title = {Sufficient {Conditions} for {Efficient} {Classical} {Simulation} of {Quantum} {Optics}},
	volume = {6},
	url = {https://link.aps.org/doi/10.1103/PhysRevX.6.021039},
	doi = {10.1103/PhysRevX.6.021039},
	number = {2},
	urldate = {2026-04-08},
	journal = {Physical Review X},
	publisher = {American Physical Society},
	author = {Rahimi-Keshari, Saleh and Ralph, Timothy C. and Caves, Carlton M.},
	month = jun,
	year = {2016},
	pages = {021039},
}

@article{mari_positive_2012,
	title = {Positive {Wigner} {Functions} {Render} {Classical} {Simulation} of {Quantum} {Computation} {Efficient}},
	volume = {109},
	url = {https://link.aps.org/doi/10.1103/PhysRevLett.109.230503},
	doi = {10.1103/PhysRevLett.109.230503},
	number = {23},
	urldate = {2026-04-08},
	journal = {Physical Review Letters},
	publisher = {American Physical Society},
	author = {Mari, A. and Eisert, J.},
	month = dec,
	year = {2012},
	pages = {230503},
}

@article{lange_electron-correlation-induced_2024,
	title = {Electron-correlation-induced nonclassicality of light from high-order harmonic generation},
	volume = {109},
	url = {https://link.aps.org/doi/10.1103/PhysRevA.109.033110},
	doi = {10.1103/PhysRevA.109.033110},
	number = {3},
	urldate = {2024-07-23},
	journal = {Physical Review A},
	author = {Lange, Christian Saugbjerg and Hansen, Thomas and Madsen, Lars Bojer},
	month = mar,
	year = {2024},
	pages = {033110},
}

@article{theidel_evidence_2024,
	title = {Evidence of the {Quantum} {Optical} {Nature} of {High}-{Harmonic} {Generation}},
	volume = {5},
	url = {https://link.aps.org/doi/10.1103/PRXQuantum.5.040319},
	doi = {10.1103/PRXQuantum.5.040319},
	number = {4},
	urldate = {2024-11-14},
	journal = {PRX Quantum},
	publisher = {American Physical Society},
	author = {Theidel, David and Cotte, Viviane and Sondenheimer, René and Shiriaeva, Viktoriia and Froidevaux, Marie and Severin, Vladislav and Merdji-Larue, Adam and Mosel, Philip and Fröhlich, Sven and Weber, Kim-Alessandro and Morgner, Uwe and Kovacev, Milutin and Biegert, Jens and Merdji, Hamed},
	month = nov,
	year = {2024},
	pages = {040319},
}

@article{theidel_observation_2025,
	title = {Observation of a displaced squeezed state in high-harmonic generation},
	volume = {7},
	issn = {2643-1564},
	url = {https://link.aps.org/doi/10.1103/6r6n-pxfp},
	doi = {10.1103/6r6n-pxfp},
	number = {3},
	urldate = {2025-10-29},
	journal = {Physical Review Research},
	author = {Theidel, David and Cotte, Viviane and Heinzel, Philip and Griguer, Houssna and Weis, Mateusz and Sondenheimer, René and Merdji, Hamed},
	month = sep,
	year = {2025},
	pages = {033223},
}

@article{stammer_high_2022,
	title = {High {Photon} {Number} {Entangled} {States} and {Coherent} {State} {Superposition} from the {Extreme} {Ultraviolet} to the {Far} {Infrared}},
	volume = {128},
	url = {https://link.aps.org/doi/10.1103/PhysRevLett.128.123603},
	doi = {10.1103/PhysRevLett.128.123603},
	number = {12},
	urldate = {2022-04-22},
	journal = {Physical Review Letters},
	author = {Stammer, Philipp and Rivera-Dean, Javier and Lamprou, Theocharis and Pisanty, Emilio and Ciappina, Marcelo F. and Tzallas, Paraskevas and Lewenstein, Maciej},
	month = mar,
	year = {2022},
	pages = {123603},
}

@article{lewenstein_generation_2021,
	title = {Generation of optical {Schrödinger} cat states in intense laser–matter interactions},
	volume = {17},
	copyright = {2021 The Author(s), under exclusive licence to Springer Nature Limited},
	issn = {1745-2481},
	url = {http://www.nature.com/articles/s41567-021-01317-w},
	doi = {10.1038/s41567-021-01317-w},
	number = {10},
	urldate = {2022-01-15},
	journal = {Nature Physics},
	author = {Lewenstein, M. and Ciappina, M. F. and Pisanty, E. and Rivera-Dean, J. and Stammer, P. and Lamprou, Th and Tzallas, P.},
	month = oct,
	year = {2021},
	pages = {1104--1108},
}

@article{gorlach_high-harmonic_2023,
	title = {High-harmonic generation driven by quantum light},
	copyright = {2023 The Author(s), under exclusive licence to Springer Nature Limited},
	issn = {1745-2481},
	url = {https://www.nature.com/articles/s41567-023-02127-y},
	doi = {10.1038/s41567-023-02127-y},
	urldate = {2023-10-09},
	journal = {Nature Physics},
	author = {Gorlach, Alexey and Tzur, Matan Even and Birk, Michael and Krüger, Michael and Rivera, Nicholas and Cohen, Oren and Kaminer, Ido},
	month = aug,
	year = {2023},
	pages = {1--8},
}

@article{antoine_attosecond_1996,
	title = {Attosecond {Pulse} {Trains} {Using} {High}--{Order} {Harmonics}},
	volume = {77},
	url = {https://link.aps.org/doi/10.1103/PhysRevLett.77.1234},
	doi = {10.1103/PhysRevLett.77.1234},
	number = {7},
	urldate = {2024-03-08},
	journal = {Physical Review Letters},
	author = {Antoine, Philippe and L'Huillier, Anne and Lewenstein, Maciej},
	month = aug,
	year = {1996},
	pages = {1234--1237},
}

@article{drescher_x-ray_2001,
	title = {X-ray {Pulses} {Approaching} the {Attosecond} {Frontier}},
	volume = {291},
	url = {https://www.science.org/doi/10.1126/science.1058561},
	doi = {10.1126/science.1058561},
	number = {5510},
	urldate = {2023-08-24},
	journal = {Science},
	author = {Drescher, Markus and Hentschel, Michael and Kienberger, Reinhard and Tempea, Gabriel and Spielmann, Christian and Reider, Georg A. and Corkum, Paul B. and Krausz, Ferenc},
	month = mar,
	year = {2001},
	pages = {1923--1927},
}

@article{paul_observation_2001,
	title = {Observation of a {Train} of {Attosecond} {Pulses} from {High} {Harmonic} {Generation}},
	volume = {292},
	url = {https://www.science.org/doi/10.1126/science.1059413},
	doi = {10.1126/science.1059413},
	number = {5522},
	urldate = {2023-09-03},
	journal = {Science},
	author = {Paul, P. M. and Toma, E. S. and Breger, P. and Mullot, G. and Augé, F. and Balcou, Ph. and Muller, H. G. and Agostini, P.},
	month = jun,
	year = {2001},
	pages = {1689--1692},
}

@article{neergaard-nielsen_generation_2006,
	title = {Generation of a {Superposition} of {Odd} {Photon} {Number} {States} for {Quantum} {Information} {Networks}},
	volume = {97},
	url = {https://link.aps.org/doi/10.1103/PhysRevLett.97.083604},
	doi = {10.1103/PhysRevLett.97.083604},
	number = {8},
	urldate = {2023-09-17},
	journal = {Physical Review Letters},
	author = {Neergaard-Nielsen, J. S. and Nielsen, B. Melholt and Hettich, C. and Mølmer, K. and Polzik, E. S.},
	month = aug,
	year = {2006},
	pages = {083604},
}

@article{ourjoumtsev_generation_2007,
	title = {Generation of optical ‘{Schrödinger} cats’ from photon number states},
	volume = {448},
	copyright = {2007 Springer Nature Limited},
	issn = {1476-4687},
	url = {https://www.nature.com/articles/nature06054},
	doi = {10.1038/nature06054},
	number = {7155},
	urldate = {2023-06-20},
	journal = {Nature},
	author = {Ourjoumtsev, Alexei and Jeong, Hyunseok and Tualle-Brouri, Rosa and Grangier, Philippe},
	month = aug,
	year = {2007},
	note = {Number: 7155},
	pages = {784--786},
}

@article{wu_squeezed_1987,
	title = {Squeezed states of light from an optical parametric oscillator},
	volume = {4},
	copyright = {© 1987 Optical Society of America},
	issn = {1520-8540},
	url = {https://opg.optica.org/josab/abstract.cfm?uri=josab-4-10-1465},
	doi = {10.1364/JOSAB.4.001465},
	number = {10},
	urldate = {2024-07-23},
	journal = {JOSA B},
	author = {Wu, Ling-An and Xiao, Min and Kimble, H. J.},
	month = oct,
	year = {1987},
	pages = {1465--1475},
}

@article{lam_optimization_1999,
	title = {Optimization and transfer of vacuum squeezing from an optical parametric oscillator},
	volume = {1},
	issn = {1464-4266},
	url = {https://dx.doi.org/10.1088/1464-4266/1/4/319},
	doi = {10.1088/1464-4266/1/4/319},
	number = {4},
	urldate = {2024-07-23},
	journal = {Journal of Optics B: Quantum and Semiclassical Optics},
	author = {Lam, P. K. and Ralph, T. C. and Buchler, B. C. and McClelland, D. E. and Bachor, H.-A. and Gao, J.},
	month = aug,
	year = {1999},
	pages = {469},
}

@article{schneider_generation_1998,
	title = {Generation of strongly squeezed continuous-wave light at 1064 nm},
	volume = {2},
	copyright = {© 1998 Optical Society of America},
	issn = {1094-4087},
	url = {https://opg.optica.org/oe/abstract.cfm?uri=oe-2-3-59},
	doi = {10.1364/OE.2.000059},
	number = {3},
	urldate = {2024-07-23},
	journal = {Optics Express},
	author = {Schneider, K. and Lang, M. and Mlynek, J. and Schiller, S.},
	month = feb,
	year = {1998},
	pages = {59--64},
}

@article{abadie_gravitational_2011,
	title = {A gravitational wave observatory operating beyond the quantum shot-noise limit},
	volume = {7},
	copyright = {2011 Springer Nature Limited},
	issn = {1745-2481},
	url = {https://www.nature.com/articles/nphys2083},
	doi = {10.1038/nphys2083},
	number = {12},
	urldate = {2023-08-21},
	journal = {Nature Physics},
	author = {{The LIGO Scientific Collaboration}},
	month = dec,
	year = {2011},
	pages = {962--965},
}

@article{xiao_precision_1987,
	title = {Precision measurement beyond the shot-noise limit},
	volume = {59},
	url = {https://link.aps.org/doi/10.1103/PhysRevLett.59.278},
	doi = {10.1103/PhysRevLett.59.278},
	number = {3},
	urldate = {2023-09-15},
	journal = {Physical Review Letters},
	author = {Xiao, Min and Wu, Ling-An and Kimble, H. J.},
	month = jul,
	year = {1987},
	pages = {278--281},
}

@article{slusher_observation_1985,
	title = {Observation of {Squeezed} {States} {Generated} by {Four}-{Wave} {Mixing} in an {Optical} {Cavity}},
	volume = {55},
	url = {https://link.aps.org/doi/10.1103/PhysRevLett.55.2409},
	doi = {10.1103/PhysRevLett.55.2409},
	number = {22},
	urldate = {2023-09-15},
	journal = {Physical Review Letters},
	author = {Slusher, R. E. and Hollberg, L. W. and Yurke, B. and Mertz, J. C. and Valley, J. F.},
	month = nov,
	year = {1985},
	pages = {2409--2412},
}

@article{walls_squeezed_1983,
	title = {Squeezed states of light},
	volume = {306},
	copyright = {1983 Springer Nature Limited},
	issn = {1476-4687},
	url = {https://www.nature.com/articles/306141a0},
	doi = {10.1038/306141a0},
	number = {5939},
	urldate = {2023-09-15},
	journal = {Nature},
	author = {Walls, D. F.},
	month = nov,
	year = {1983},
	pages = {141--146},
}

@article{horodecki_quantum_2009,
	title = {Quantum entanglement},
	volume = {81},
	url = {https://link.aps.org/doi/10.1103/RevModPhys.81.865},
	doi = {10.1103/RevModPhys.81.865},
	number = {2},
	urldate = {2022-11-03},
	journal = {Reviews of Modern Physics},
	publisher = {American Physical Society},
	author = {Horodecki, Ryszard and Horodecki, Paweł and Horodecki, Michał and Horodecki, Karol},
	month = jun,
	year = {2009},
	pages = {865--942},
}

@article{usenko_continuous-variable_2026,
	title = {Continuous-variable quantum communication},
	volume = {98},
	url = {https://link.aps.org/doi/10.1103/mgj7-t6d3},
	doi = {10.1103/mgj7-t6d3},
	number = {1},
	urldate = {2026-04-08},
	journal = {Reviews of Modern Physics},
	publisher = {American Physical Society},
	author = {Usenko, Vladyslav C. and Acín, Antonio and Alléaume, Romain and Andersen, Ulrik L. and Diamanti, Eleni and Gehring, Tobias and Hajomer, Adnan A. E. and Kanitschar, Florian and Pacher, Christoph and Pirandola, Stefano and Pruneri, Valerio},
	month = mar,
	year = {2026},
	pages = {015003},
}

@article{rivera-dean_attosecond_2026,
	title = {Attosecond quantum optical interferometry},
	volume = {89},
	issn = {0034-4885},
	url = {https://doi.org/10.1088/1361-6633/ae5847},
	doi = {10.1088/1361-6633/ae5847},
	number = {4},
	urldate = {2026-05-25},
	journal = {Reports on Progress in Physics},
	author = {Rivera-Dean, Javier and Petrovic, Lidija and Lewenstein, Maciej and Stammer, Philipp},
	month = apr,
	year = {2026},
	pages = {047901},
}

@misc{stammer_fluctuation-induced_2026,
	title = {Fluctuation-induced symmetry breaking in high harmonic generation for bicircular quantum light},
	url = {http://arxiv.org/abs/2603.24377},
	doi = {10.48550/arXiv.2603.24377},
	urldate = {2026-04-08},
	publisher = {arXiv},
	author = {Stammer, Philipp and Granados, Camilo and Rivera-Dean, Javier},
	month = apr,
	year = {2026},
	note = {arXiv:2603.24377 [quant-ph]},
}

@article{stammer_quantum_2023,
	title = {Quantum {Electrodynamics} of {Intense} {Laser}-{Matter} {Interactions}: {A} {Tool} for {Quantum} {State} {Engineering}},
	volume = {4},
	shorttitle = {Quantum {Electrodynamics} of {Intense} {Laser}-{Matter} {Interactions}},
	url = {https://link.aps.org/doi/10.1103/PRXQuantum.4.010201},
	doi = {10.1103/PRXQuantum.4.010201},
	number = {1},
	urldate = {2023-01-25},
	journal = {PRX Quantum},
	publisher = {American Physical Society},
	author = {Stammer, Philipp and Rivera-Dean, Javier and Maxwell, Andrew and Lamprou, Theocharis and Ordóñez, Andrés and Ciappina, Marcelo F. and Tzallas, Paraskevas and Lewenstein, Maciej},
	month = jan,
	year = {2023},
	pages = {010201},
}

@article{lemieux_photon_2025,
	title = {Photon bunching in high-harmonic emission controlled by quantum light},
	volume = {19},
	copyright = {2025 Crown},
	issn = {1749-4893},
	url = {https://www.nature.com/articles/s41566-025-01673-6},
	doi = {10.1038/s41566-025-01673-6},
	number = {7},
	urldate = {2025-12-01},
	journal = {Nature Photonics},
	author = {Lemieux, Samuel and Jalil, Sohail A. and Purschke, David N. and Boroumand, Neda and Hammond, T. J. and Villeneuve, David and Naumov, Andrei and Brabec, Thomas and Vampa, Giulio},
	month = jul,
	year = {2025},
	pages = {767--771},
}

@misc{tzur_measuring_2025,
	title = {Measuring and controlling the birth of quantum attosecond pulses},
	url = {http://arxiv.org/abs/2502.09427},
	doi = {10.48550/arXiv.2502.09427},
	urldate = {2025-02-24},
	publisher = {arXiv},
	author = {Tzur, Matan Even and Mor, Chen and Yaffe, Noa and Birk, Michael and Rasputnyi, Andrei and Kneller, Omer and Nisim, Ido and Kaminer, Ido and Krüger, Michael and Dudovich, Nirit and Cohen, Oren},
	month = feb,
	year = {2025},
	note = {arXiv:2502.09427 [physics]},
}

@article{heimerl_multiphoton_2024,
	title = {Multiphoton electron emission with non-classical light},
	volume = {20},
	copyright = {2024 The Author(s), under exclusive licence to Springer Nature Limited},
	issn = {1745-2481},
	url = {https://www.nature.com/articles/s41567-024-02472-6},
	doi = {10.1038/s41567-024-02472-6},
	number = {6},
	urldate = {2025-08-12},
	journal = {Nature Physics},
	author = {Heimerl, Jonas and Mikhaylov, Alexander and Meier, Stefan and Höllerer, Henrick and Kaminer, Ido and Chekhova, Maria and Hommelhoff, Peter},
	month = jun,
	year = {2024},
	pages = {945--950},
}

@article{heimerl_driving_2025,
	title = {Quantum light drives electrons strongly at metal needle tips},
	volume = {21},
	copyright = {2025 The Author(s)},
	issn = {1745-2481},
	url = {https://www.nature.com/articles/s41567-025-03087-1},
	doi = {10.1038/s41567-025-03087-1},
	number = {12},
	urldate = {2026-04-10},
	journal = {Nature Physics},
	author = {Heimerl, Jonas and Rasputnyi, Andrei and Pölloth, Jonathan and Meier, Stefan and Chekhova, Maria and Hommelhoff, Peter},
	month = dec,
	year = {2025},
	pages = {1899--1904},
}

@article{spasibko_multiphoton_2017,
	title = {Multiphoton {Effects} {Enhanced} due to {Ultrafast} {Photon}-{Number} {Fluctuations}},
	volume = {119},
	url = {https://link.aps.org/doi/10.1103/PhysRevLett.119.223603},
	doi = {10.1103/PhysRevLett.119.223603},
	number = {22},
	urldate = {2025-02-06},
	journal = {Physical Review Letters},
	publisher = {American Physical Society},
	author = {Spasibko, Kirill Yu. and Kopylov, Denis A. and Krutyanskiy, Victor L. and Murzina, Tatiana V. and Leuchs, Gerd and Chekhova, Maria V.},
	month = nov,
	year = {2017},
	pages = {223603},
}

@article{manceau_indefinite-mean_2019,
	title = {Indefinite-{Mean} {Pareto} {Photon} {Distribution} from {Amplified} {Quantum} {Noise}},
	volume = {123},
	url = {https://link.aps.org/doi/10.1103/PhysRevLett.123.123606},
	doi = {10.1103/PhysRevLett.123.123606},
	number = {12},
	urldate = {2025-02-06},
	journal = {Physical Review Letters},
	publisher = {American Physical Society},
	author = {Manceau, Mathieu and Spasibko, Kirill Yu. and Leuchs, Gerd and Filip, Radim and Chekhova, Maria V.},
	month = sep,
	year = {2019},
	pages = {123606},
}

@article{rasputnyi_high_2024,
	title = {High-harmonic generation by a bright squeezed vacuum},
	copyright = {2024 The Author(s)},
	issn = {1745-2481},
	url = {https://www.nature.com/articles/s41567-024-02659-x},
	doi = {10.1038/s41567-024-02659-x},
	urldate = {2024-11-11},
	journal = {Nature Physics},
	author = {Rasputnyi, Andrei and Chen, Zhaopin and Birk, Michael and Cohen, Oren and Kaminer, Ido and Krüger, Michael and Seletskiy, Denis and Chekhova, Maria and Tani, Francesco},
	month = oct,
	year = {2024},
}

@article{stammer_squeezing_2023,
	title = {Entanglement and {Squeezing} of the {Optical} {Field} {Modes} in {High} {Harmonic} {Generation}},
	volume = {132},
	url = {https://link.aps.org/doi/10.1103/PhysRevLett.132.143603},
	doi = {10.1103/PhysRevLett.132.143603},
	number = {14},
	urldate = {2024-05-14},
	journal = {Physical Review Letters},
	author = {Stammer, Philipp and Rivera-Dean, Javier and Maxwell, Andrew S. and Lamprou, Theocharis and Argüello-Luengo, Javier and Tzallas, Paraskevas and Ciappina, Marcelo F. and Lewenstein, Maciej},
	month = apr,
	year = {2024},
	pages = {143603},
}

@article{lewenstein_theory_1994,
	title = {Theory of high-harmonic generation by low-frequency laser fields},
	volume = {49},
	url = {https://link.aps.org/doi/10.1103/PhysRevA.49.2117},
	doi = {10.1103/PhysRevA.49.2117},
	number = {3},
	urldate = {2022-06-16},
	journal = {Physical Review A},
	author = {Lewenstein, M. and Balcou, Ph. and Ivanov, M. Yu. and L’Huillier, Anne and Corkum, P. B.},
	month = mar,
	year = {1994},
	pages = {2117--2132},
}

@article{amini_symphony_2019,
	title = {Symphony on strong field approximation},
	volume = {82},
	issn = {0034-4885},
	url = {https://doi.org/10.1088/1361-6633/ab2bb1},
	doi = {10.1088/1361-6633/ab2bb1},
	number = {11},
	urldate = {2021-12-01},
	journal = {Reports on Progress in Physics},
	author = {Amini, Kasra and Biegert, Jens and Calegari, Francesca and Chacón, Alexis and Ciappina, Marcelo F. and Dauphin, Alexandre and Efimov, Dmitry K. and Faria, Carla Figueira de Morisson and Giergiel, Krzysztof and Gniewek, Piotr and Landsman, Alexandra S. and Lesiuk, Micha{\textbackslash}l and Mandrysz, Micha{\textbackslash}l and Maxwell, Andrew S. and Moszyński, Robert and Ortmann, Lisa and Pérez-Hernández, Jose Antonio and Picón, Antonio and Pisanty, Emilio and Prauzner-Bechcicki, Jakub and Sacha, Krzysztof and Suárez, Noslen and Zaïr, Amelle and Zakrzewski, Jakub and Lewenstein, Maciej},
	month = oct,
	year = {2019},
	pages = {116001},
}

@article{ourjoumtsev_generating_2006,
	title = {Generating {Optical} {Schrödinger} {Kittens} for {Quantum} {Information} {Processing}},
	volume = {312},
	url = {https://www.science.org/doi/10.1126/science.1122858},
	doi = {10.1126/science.1122858},
	number = {5770},
	urldate = {2023-06-20},
	journal = {Science},
	author = {Ourjoumtsev, Alexei and Tualle-Brouri, Rosa and Laurat, Julien and Grangier, Philippe},
	month = apr,
	year = {2006},
	pages = {83--86},
}

@article{rivera-dean_role_2024,
	title = {Role of short and long trajectories on the quantum-optical state after high-order harmonic generation},
	volume = {110},
	url = {https://link.aps.org/doi/10.1103/PhysRevA.110.063704},
	doi = {10.1103/PhysRevA.110.063704},
	number = {6},
	urldate = {2025-01-06},
	journal = {Physical Review A},
	author = {Rivera-Dean, Javier},
	month = dec,
	year = {2024},
	pages = {063704},
}

\newpage
\clearpage
\appendix
\onecolumngrid
\begin{center}
	\large \textbf{\textsc{Supplementary Material}}
\end{center}

\section{Single quadrature expansion of squeezed states of light}\label{Sec:SM:1D:model}
Throughout this work, single-mode squeezed states of light are represented using a single-quadrature coherent state expansion, i.e., as a superposition of coherent states aligned along a given optical quadrature and weighted by a real-valued Gaussian probability amplitude.~In this section, we demonstrate that such kind of superpositions exactly represent squeezed vacuum states aligned along the same optical quadrature for squeezing parameters within the regime of interest for this work.~The proof presented here follows a top-down approach: starting from the specific expression of the state, we demonstrate its equivalence to a squeezed state.~The corresponding bottom-up approach will be presented elsewhere.

We begin by introducing the proposed single-quadrature description, which is given by
\begin{equation}
	\ket{\phi_{\text{1D}}}
	= \sqrt{\dfrac{(1+\sigma)^{\frac12}}{\pi \sigma}}
	\int_{\mathbbm{R}}\dd \alpha
	\exp[-\dfrac{\alpha^2}{\sigma}]
	\ket{\alpha},
\end{equation}
where $\sigma = e^{\bar{r}} \cosh(\bar{r})$.~This state is a Gaussian state,~as can be readily verified by computing its Wigner function.~The latter is given by~\cite{royer_wigner_1977}
\begin{equation}
	W_{\text{1D}}(\beta)
	= \dfrac{2}{\pi}
	\mel{\phi_{\text{1D}}}{\hat{D}(-\beta)\hat{\Pi}\hat{D}(\beta)}{\phi_{\text{1D}}}
	= \dfrac{2}{\pi}
	\exp[-\dfrac{\beta_x^2}{(1+\sigma)/2}
	-\dfrac{\beta_y^2}{1/[2(1+\sigma)]}],
\end{equation}
from which we can identify the variances along the two optical quadratures as
\begin{equation}
	\Big[\Delta X_1^{(\text{1D})}\Big]^2
	= \dfrac{1+\sigma}{2},
	\quad \quad
	\Big[\Delta X_2^{(\text{1D})}\Big]^2
	= \dfrac{1}{2(1+\sigma)}.
\end{equation}
It is worth noting that these variances saturate the Heisenberg uncertainty relation, $\Delta X_1^2 \cdot \Delta X_2^2 = 1/4$.

Since the state is Gaussian, it is fully characterized by its covariance matrix and first-order moments.~The latter vanishes in this specific case, as can be seen from the fact that the Wigner function is centered at the origin.~Therefore, two pure Gaussian states with the same covariance matrix and vanishing first moments are equivalent up to an irrelevant global phase.~For a squeezed-vacuum state stretched along the $X_1$ optical quadrature, with squeezing strength $r$, the corresponding variances are
\begin{equation}
	\Big[\Delta X_1^{(\text{BSV})}\Big]^2
	= \dfrac{e^{2r}}{2},
	\quad \quad
	\Big[\Delta X_2^{(\text{BSV})}\Big]^2
	= \dfrac{e^{-2r}}{2}.
\end{equation}
Equating the corresponding variances of the two states yields the relation between $r_1$ and $r_2$ for which they are equivalent,
\begin{equation}\label{Eq:SM:rel:r1:r2}
	1+e^{\bar{r}}\cosh(\bar{r}) = e^{2r}
	\implies r = \dfrac{1}{2} \ln[1+e^{\bar{r}}\cosh(\bar{r})]
	\iff \bar{r} = \dfrac{1}{2}\ln[-3+2e^{2r}].
\end{equation}
The latter expression is real-valued provided that $r \geq  \frac12\ln(3/2) \simeq 0.203$, which is well within the range of squeezing parameters considered in this work ($r\sim 10$).

Introducing the relation in Eq.~\eqref{Eq:SM:rel:r1:r2} inside Eq.~\eqref{Eq:Fid:1D}, we arrive at
\begin{equation}\label{Eq:SM:Fid:subs}
	\mathcal{F}
	= \sqrt{\dfrac{2 e^{r}\sqrt{1+ \sqrt{2e^{2r}-3} \cosh(\tfrac12 \ln[2e^{2r}-3])}}{e^{2r}+1+ \sqrt{2e^{2r}-3} \cosh(\tfrac12 \ln[2e^{2r}-3])}},
\end{equation}
and taking into account that $\cosh[(\ln x)/2] = (x+1)/(2\sqrt{x})$, one therefore gets
\begin{equation}
	1+ \sqrt{2e^{2r}-3} \cosh(\tfrac12 \ln[2e^{2r}-3])
	= e^{2r},
\end{equation}
which substituted in Eq.~\eqref{Eq:SM:Fid:subs} yields $\mathcal{F} = 1$, valid within the domain of Eq.~\eqref{Eq:SM:rel:r1:r2}.

\section{Influence of the $\varphi_2(\alpha)$ phase on the properties of the state}\label{Sec:SM:Phase:influence}
In this section of the Supplementary Material, we characterize the influence of the $\varphi_2(\alpha)$ phase on the quantum optical properties of the state.~We begin by considering the single-mode collective properties of the state, namely, by writing the quantum optical state after the HHG interaction as
\begin{equation}
	\relaxket{\Tilde{\Phi}_{\text{even}}(t)}
		= \dfrac{1}{\sqrt{\mathcal{N}}}
			\! \int_{\mathbbm{R}}\!\! \dd \alpha \ e^{iB_2\alpha^2}
				 \tilde{c}(\alpha)
					\hat{D}_{\text{even}}(\bar\chi \alpha) \ket{0}.	
\end{equation}
Following a similar analysis to that presented in the main text, we begin by confirming the Gaussian nature of the state through the evaluation of its wavefunction in an arbitrary quadrature basis, i.e., $\hat{X}_{\theta} = \hat{X}_1\cos(\theta) + \hat{X}_2 \sin(\theta)$. We then obtain
\begin{equation}
	\Tilde{\Phi}_{\text{even}}(X_\theta)
		= \langle X_1\vert e^{i\theta \hat{A}^\dagger \hat{A}}\vert \Tilde{\Phi}_{\text{even}}(t)\rangle
		= \dfrac{1}{\sqrt{\pi \mathcal{N}}}
				e^{-C \frac{\hat{X}^2_1}{2}},
		\ \text{with} \
		C = \dfrac{[B_2 \sigma + \bar{\chi}^2\sigma \cos(\theta)\sin(\theta)] + i[1+\bar{\chi}^2\sigma \sin[2](\theta)]}{[B_2 \sigma - \bar{\chi}^2\sigma \cos(\theta)\sin(\theta)] + i[1+\bar{\chi}^2\sigma \sin[2](\theta)]}.
\end{equation}
Since $\text{Re}[C] \geq 0$, the wavefunction remains Gaussian for all optical quadratures.~This is expected, as conditions (1) and (2) discussed in Sec.~\ref{Sec:Conds:Gauss} remain satisfied even in the presence of the phase $\varphi_2(\alpha) = B_2\alpha^2$.

In this case, however, the evaluation of the covariance matrix becomes slightly more complicated, as the presence of $\varphi_2(\alpha)$ breaks its a priori diagonal structure.~This can be seen from the expectation values of the second-order products of the quadrature operators
\begin{align}
	&\langle \hat{X}_1^2\rangle
		= \dfrac12
			+\dfrac{\bar{\chi}^2}{2 \mathcal{N}}
			\dfrac{\pi \sigma^2 (1+\bar{\chi}^2\sigma)}{[(1+\bar{\chi}^2 \sigma + B_2^2\sigma^2)]^{3/2}},
	\\&
	\langle \hat{X}_2^2\rangle
		= \dfrac12
			-\dfrac{\bar{\chi}^2}{2 \mathcal{N}}
			\dfrac{\pi \sigma^2}{[(1+\bar{\chi}^2 \sigma + B_2^2\sigma^2)]^{3/2}},
	\\&
	\langle \hat{X}_1\hat{X}_2\rangle
		= -\dfrac{i}{2}
			-\dfrac{\bar{\chi}^2}{2 \mathcal{N}}
			\dfrac{\pi B_2 \sigma^3}{[(1+\bar{\chi}^2 \sigma + B_2^2\sigma^2)]^{3/2}}.
\end{align} 
To incorporate many-body effects, we follow the same procedure as in the main text, substituting $\chi_q(\alpha) \to N \chi_q(\alpha)$, with $N$ the number of atoms, for all harmonic modes except for the fundamental mode. For the latter, to account for the increasing role of $\varphi_2(\alpha)$ on the quantum optical state properties, we consider the parametrization $\chi_2(\alpha) = [(1-\varrho_x A_2) + i \varrho_y B_2]\alpha$, treating both $\varrho_x$ and $\varrho_y$ as dynamical parameters.~It is worth noting that while both real and imaginary contributions to $\chi_2(\alpha)$ originate from the same microscopic dipole dynamics rooted in the HHG dipole $\langle \hat{d}(t)\rangle$, propagation effects, phase matching conditions, and electron backaction~\cite{rivera-dean_role_2024} can renormalize them differently.~However, to ensure physical consistency, i.e., no net photon gain in the driving mode, we choose the values of $\varrho_y$ satisfying
\begin{equation}
	(1-\varrho_x A_2)^2 + \varrho_y^2 B_2^2 \leq 1
		\implies
	\kappa_y \leq B_2^{-1}\sqrt{1-(1-\varrho_x A_2)^2},
\end{equation}
while sampling values of $\varrho_x$ such that $\abs{\bar{\chi}}\leq 1$, ensuring that the total emitted field does not exceed the input field amplitude.

\begin{figure}[h!]
	\centering
	\includegraphics[width=1\textwidth]{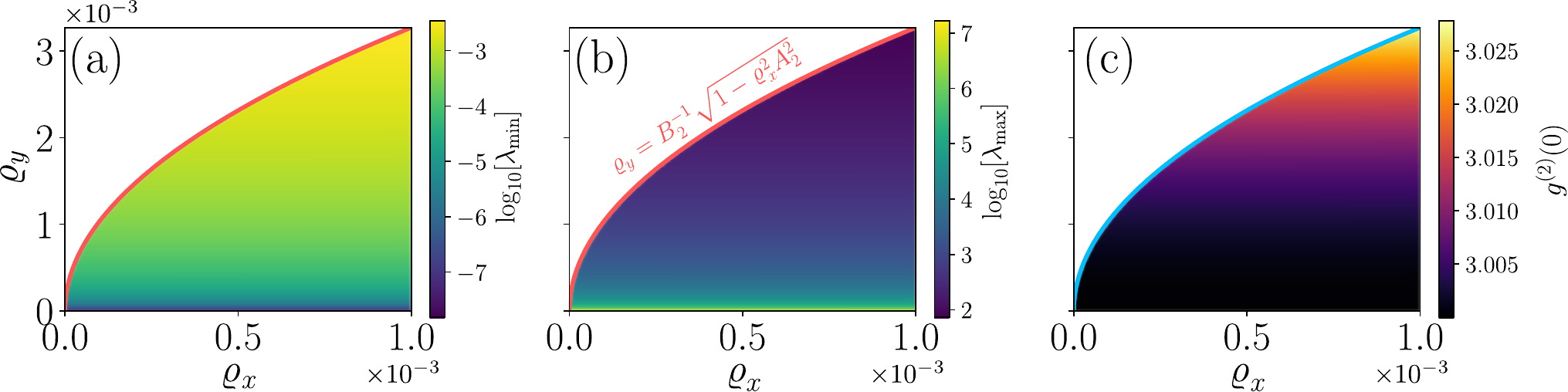}
	\caption{Analysis of single-mode properties of the state when using a 2D parametrization of the state.~(a,b) Minimum and maximum eigenvalues of the covariance matrix, respectively, represented in logarithmic scale.~(c) Second-order autocorrelation function.~Calculations have been performed for xenon atoms ($I_p\simeq 0.446$ a.u.), using a driving field frequency $\omega = 0.057$ a.u., a coherent field strength $\varepsilon_\omega = 0.053$ a.u., and a squeezing parameter $\bar{r}=9$ corresponding to a squeezing intensity $I_{\text{squ}} \sim 10^{-7}$ a.u.~for $\kappa = 5 \times 10^{-8}$ a.u.}
	\label{Fig:SM:SM}
\end{figure}

Figure~\ref{Fig:SM:SM} displays the results of this analysis.~Panels~(a) and (b) show, on a logarithmic scale, the minimum and maximum values of the eigenvalues of the covariance matrix, respectively, while panel (c) presents the second-order autocorrelation function of the state.~As observed in the first two panels, the global state still exhibits squeezing-like signatures, although increasing values of $\varrho_y$ result in a noticeable reduction of the contrast between the squeezed and anti-squeezed quadratures.~Nevertheless, for all parameter values considered, the photon number statistics remain super-Poissonian and close to those of BSV sources, with $g^{(2)}(0)\simeq 3$ across the explored range of $\varrho_x$ and $\varrho_y$.

Focusing now on bipartite correlations of the state, obtained by writing 
\begin{equation}
	\relaxket{\Tilde{\Phi}_{\text{even}}(t)}
		= \dfrac{1}{\sqrt{\mathcal{N}}}
			\! \int_{\mathbbm{R}}\!\! \dd \alpha \ e^{iB_2\alpha^2}
				\tilde{c}(\alpha)
					\ket{\chi_{\mathsf{A}}\alpha, \chi_{\mathsf{B}}\alpha},	
\end{equation}
for an arbitrary bipartition into subsystems $\mathsf{A}$ and $\mathsf{B}$ of the even harmonic modes satisfying $\mathsf{A}\cup \mathsf{B} = \{q=2n:n\in \mathbbm{N}\}$, the second-order quadrature moments read
\begin{align}
	&\langle \hat{X}_{1,\mathsf{A}}^2\rangle
		= \dfrac12
		+\dfrac{\bar{\chi}_{\mathsf{A}}^2}{2 \mathcal{N}}
		\dfrac{\pi \sigma^2 			[1+(\bar{\chi}^2_{\mathsf{A}}+\bar{\chi}^2_{\mathsf{B}})\sigma]}{[(1+(\bar{\chi}^2_{\mathsf{A}}+\bar{\chi}^2_{\mathsf{B}}) \sigma + B_2^2\sigma^2)]^{3/2}},
	\\&
	\langle \hat{X}_{2,\mathsf{A}}^2\rangle
		= \dfrac12
		-\dfrac{\bar{\chi}_{\mathsf{A}}^2}{2 \mathcal{N}}
		\dfrac{\pi \sigma^2}{[(1+(\bar{\chi}^2_{\mathsf{A}}+\bar{\chi}^2_{\mathsf{B}}) \sigma + B_2^2\sigma^2)]^{3/2}},
	\\&
	\langle \hat{X}_{1,\mathsf{A}}\hat{X}_{2,\mathsf{A}}\rangle
		= -\dfrac{i}{2}
		-\dfrac{\bar{\chi}^2_{\mathsf{A}}}{2 \mathcal{N}}
		\dfrac{\pi B_2 \sigma^3}{[(1+(\bar{\chi}^2_{\mathsf{A}}+\bar{\chi}^2_{\mathsf{B}})  \sigma + B_2^2\sigma^2)]^{3/2}},
	\\&
	\langle \hat{X}_{1,\mathsf{A}}\hat{X}_{1,\mathsf{B}}\rangle
		= \dfrac{\bar{\chi}_{\mathsf{A}}\bar{\chi}_{\mathsf{B}}}{2 \mathcal{N}}
		\dfrac{\pi \sigma^2 [1+(\bar{\chi}^2_{\mathsf{A}}+\bar{\chi}^2_{\mathsf{B}})\sigma]}{[(1+(\bar{\chi}^2_{\mathsf{A}}+\bar{\chi}^2_{\mathsf{B}}) \sigma + B_2^2\sigma^2)]^{3/2}},
	\\&\langle \hat{X}_{2,\mathsf{A}}\hat{X}_{2,\mathsf{B}}\rangle
		= -\dfrac{\bar{\chi}_{\mathsf{A}}\bar{\chi}_{\mathsf{B}}}{2 \mathcal{N}}
		\dfrac{\pi \sigma^2}{[(1+(\bar{\chi}^2_{\mathsf{A}}+\bar{\chi}^2_{\mathsf{B}}) \sigma + B_2^2\sigma^2)]^{3/2}},
	\\&\langle \hat{X}_{1,\mathsf{A}}\hat{X}_{2,\mathsf{B}}\rangle
		= -\dfrac{\bar{\chi}_{\mathsf{A}}\bar{\chi}_{\mathsf{B}}}{2 \mathcal{N}}
		\dfrac{B_2\pi \sigma^3}{[(1+(\bar{\chi}^2_{\mathsf{A}}+\bar{\chi}^2_{\mathsf{B}}) \sigma + B_2^2\sigma^2)]^{3/2}}.
\end{align}
All remaining elements follow from symmetry under the exchange $\mathsf{A}\leftrightarrow\mathsf{B}$, together with the block structure of the covariance matrix.

\begin{figure}
	\centering
	\includegraphics[width=1\textwidth]{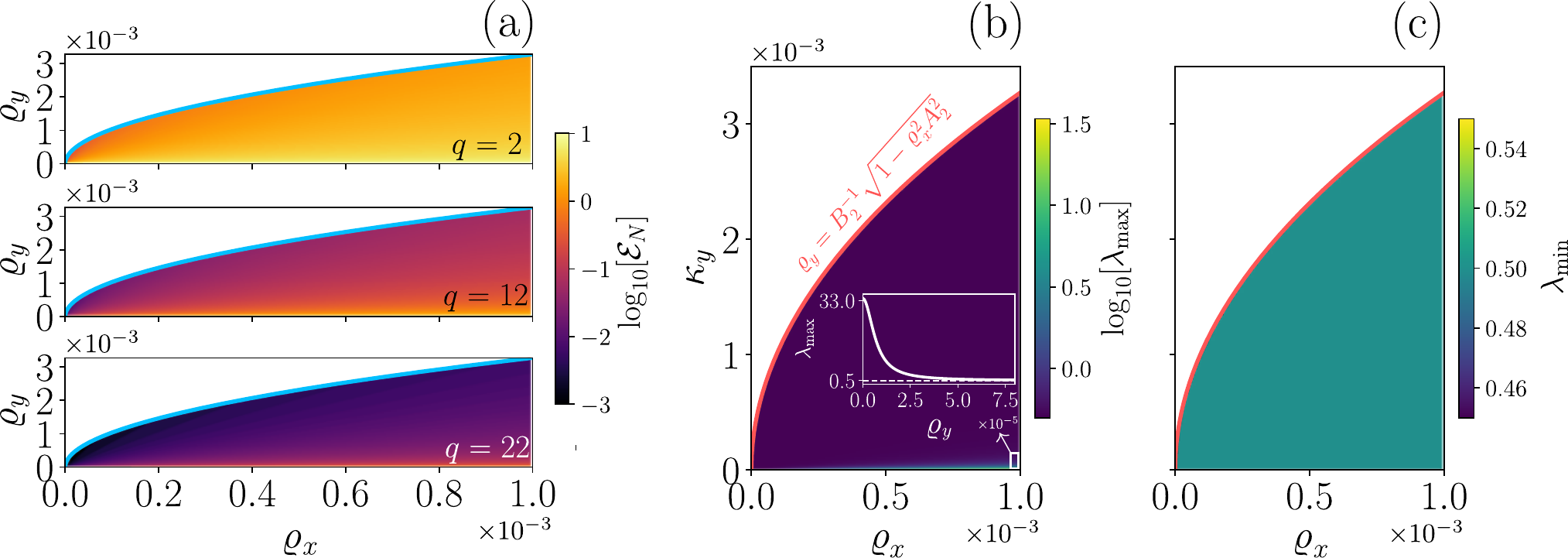}
	\caption{Analysis of two-mode properties of the state with bipartitions $\mathsf{A} = \{q\}$ and $\mathsf{B}= \bar{\mathsf{A}}$, when using a 2D parametrization of the state.~(a,b) Logarithmic negativity (in logarithmic scale) for three different harmonic orders corresponding to the $2\omega$ driving field (upper panel), a plateau harmonic (middle panel) and a cutoff harmonic (lower panel).~(b,c) Minimum and maximum eigenvalues of the covariance matrix of subsystem $\mathsf{A}$ for the 10th harmonic order, respectively, represented in logarithmic scale.~Calculations have been performed for xenon atoms ($I_p\simeq 0.446$ a.u.), using a driving field frequency $\omega = 0.057$ a.u., a coherent field strength $\varepsilon_\omega = 0.053$ a.u., and a squeezing parameter $\bar{r}=9$ corresponding to a squeezing intensity $I_{\text{squ}} \sim 10^{-7}$ a.u.~for $\kappa = 5 \times 10^{-8}$ a.u.}
	\label{Fig:TM:SM}
\end{figure}

Figure~\ref{Fig:TM:SM}~(a) shows the logarithmic negativity for different harmonic modes, where subsystem $\mathsf{A}:=\{q\}$ and $\mathsf{B}:=\{q'\neq q : q\in \text{even}\}$.~From top to bottom, the panels correspond to the $2\omega$ driving field, a harmonic in the plateau region, and a harmonic in the cutoff region, respectively.~As observed, the strongest correlations still occur between the driving mode and the remaining modes, decreasing towards the plateau region and becoming negligible in the cutoff region. Similarly to the single-mode case, increasing $\varrho_y$ leads to a systematic reduction of the entanglement by approximately one order of magnitude across all cases, for the largest $\varrho_y$ compatible with our constraints.

Panels~(b) and (c) of Fig.~\ref{Fig:TM:SM} show the maximum and minimum eigenvalues of the reduced covariance matrix of subsystem $\mathsf{A}$ for the 10th harmonic order.~The minimum eigenvalue saturates the vacuum limit independently of $\varrho_x$ and $\varrho_y$, while the maximum eigenvalue exhibits strong anti-squeezing for $\varrho_y\simeq 0$, which is progressively suppressed as $\varrho_y$ increases, eventually approaching the vacuum.~Comparison with recent experimental results using similar interferometric configurations~\cite{tzur_attosecond-resolved_2025} shows that the reconstructed Husimi-like functions exhibit clear anti-squeezing along one quadrature, therefore suggesting that in the regime $\varrho_y \simeq 0$ provides the best agreement with experimental observations, and is accordingly the choice adopted in the main text.

\section{Numerical optimization of the Duan-Simon inequality violation}\label{Sec:SM:DSI:opt}
In this section, we briefly describe how the optimization of the Duan-Simon inequality violation defined in Eq.~\eqref{Eq:DSI} was performed in practice.~This expression does not correspond to a single inequality, but rather to a family of inequalities parametrized by $a \in \mathbbm{R}$.~Its optimization therefore relies on finding an optimal value of $\mathsf{a}$ that maximizes the observed violation, which in general depends on the choice of bipartitions $\mathsf{A}$ and $\mathsf{B}$, as well on the atomic system under consideration.~In this work, the optimization was carried out as follows:
\begin{itemize}
	\item Having chosen a bipartition $\mathsf{A}$ and $\mathsf{B}$, we define two sets of values for $a$, each containing around $2.5\times 10^4$ elements.~These sets are constructed by uniformly sampling the exponents in the range $[-8,8]$, yielding one logarithmically spaced grid spanning $[10^{-8},10^8]$, and a symmetric negative counterpart spanning $[-10^{8},-10^{-8}]$. This logarithmic sampling avoids the divergence at $a=0$ while ensuring high resolution near the origin and extensive coverage across multiple orders of magnitude.~Minor adjustments to the grid density or range were made in cases where the results for the different atomic species exhibit non-monotonic behavior as a function of $\varrho$, indicating suboptimal convergence.
	\item We evaluate the DSI violation at each element of both sets and retain the maximum value found in each, which we refer to as $\pm a_0$.
	\item Setting $\pm a_0$ as initial values, we perform a local optimization for both positive and negative branches using the Nelder-Mead algorithm, yielding the optimal values $\pm a^*$.
	\item Finally, we take the maximum between $\text{DSI}[a^*]$ and $\text{DSI}[-a^*]$ as the reported value.
\end{itemize}

\section{Analysis of single- and multimode (time-zero) second-order correlation functions}\label{Sec:SM:g2}
An alternative method of witnessing entanglement, recently used in the context of strong-field physics~\cite{theidel_evidence_2024}, relies on the violation of a Cauchy-Schwarz inequality written in terms of time-zero second-order correlation functions.~More specifically, one has that
\begin{equation}
	\big[ g_{\mathsf{A},\mathsf{B}}^{(2)}\big]^2
		 \leq g_{\mathsf{A},\mathsf{A}}^{(2)} g_{\mathsf{B},\mathsf{B}}^{(2)},
	\ \text{with} \
	g_{i,j}^{(2)}
		= \dfrac{\langle \hat{a}^\dagger_i \hat{a}^\dagger_j \hat{a}_i \hat{a}_j \rangle}{\langle \hat{a}^\dagger_i \hat{a}_i \rangle\langle \hat{a}^\dagger_j \hat{a}_j \rangle},
\end{equation} 
with violation of the inequality being a sufficient, but no necessary, condition for entanglement.~Interestingly, given an arbitrary bipartition $\mathsf{A}$ and $\mathsf{B}$ of the even harmonic set satisfying $\mathsf{A}\cup \mathsf{B} = \{q=2n:n\in \mathbbm{N}\}$, the second-order autocorrelation functions of each subsystem read
\begin{equation}
	\begin{aligned}
		&\langle \hat{a}_\mathsf{A}^{\dagger 2} \hat{a}_{\mathsf{A}}^2 \rangle
		= \int_{\mathbbm{R}} \dd \alpha \int_{\mathbbm{R}}
		\dd \beta
		\ c^*(\beta)c(\alpha) \chi_{\mathsf{A}}^4 \alpha^2 \beta^2
		\braket{\chi_{\mathsf{A}}\beta}{\chi_{\mathsf{A}}\alpha}
		\braket{\chi_{\mathsf{B}}\beta}{\chi_{\mathsf{B}}\alpha}
		= K\chi_{\mathsf{A}}^4, 
		\\& \langle \hat{a}_\mathsf{B}^{\dagger 2} \hat{a}_{\mathsf{B}}^2 \rangle 
		= K\chi_{\mathsf{B}}^4 ,
	\end{aligned}
\end{equation}
while the cross-correlation between both subsystems gives
\begin{equation}
	\begin{aligned}
		&\langle \hat{a}_\mathsf{A}^{\dagger} \hat{a}_\mathsf{B}^{\dagger} \hat{a}_{\mathsf{A}} \hat{a}_{\mathsf{B}}\rangle
		= \int_{\mathbbm{R}} \dd \alpha \int_{\mathbbm{R}}
		\dd \beta
		\ c^*(\beta)c(\alpha) \chi_{\mathsf{A}}^2 \chi_{\mathsf{B}}^2 \alpha^2 \beta^2
		\braket{\chi_{\mathsf{A}}\beta}{\chi_{\mathsf{A}}\alpha}
		\braket{\chi_{\mathsf{B}}\beta}{\chi_{\mathsf{B}}\alpha}
		= K\chi_{\mathsf{A}}^2 \chi_{\mathsf{B}}^2 ,
	\end{aligned}
\end{equation}
which naturally results in $[g_{\mathsf{A},\mathsf{B}}^{(2)}]^2
= g_{\mathsf{A},\mathsf{A}}^{(2)} g_{\mathsf{B},\mathsf{B}}^{(2)}$, thereby saturating the Cauchy-Schwartz inequality, and precluding entanglement certification through this approach for the present configuration~\cite{rivera-dean_attosecond_2026}.

Finally, when considering a collective mode analysis, the second-order autocorrelation function of the global mode reads
\begin{equation}
	g^{(2)}(0)
		= 3 
		+ \bigg[
				\dfrac{1}{\bar{\chi}^2\sigma} + \dfrac{1}{\bar{\chi}^4\sigma^2}
			\bigg].
\end{equation}
Under low depletion conditions ($\bar{\chi}\lesssim 1$), the behavior of this expression is primarily determined by $\sigma$, which satisfies $\sigma \gg 1$ in our case, yielding $g^{(2)}(0)\simeq 3$.

\section{About the Gaussian state regime}\label{Sec:SM:Gaussian:valid}
As stated in Sec.~\ref{Sec:Conds:Gauss}, a sufficient condition for the quantum optical state after the HHG interaction to remain Gaussian relies on the affine functional dependence of $\chi(\alpha)$ which, as discussed in Sec.~\ref{Sec:Brkdwn:gaussian}, does not hold when the squeezing strength becomes particularly large. In what follows, we discuss the conditions that $\bar{r}$, and thereby $r$, needs to satisfy for the state to remain Gaussian.

In our bichromatic context, the Gaussian condition is in general imposed by the validity of the Taylor expansions considered Eqs.~\eqref{Eq:field:amp:odd} and \eqref{Eq:field:amp:even}, which determine the harmonic conversion efficiency of both even and odd harmonic orders. Thus, when considering the next higher-order terms, one can generally write
\begin{equation}\label{Eq:SM:chi:higher:order}
	\chi_q(\alpha) =
	\left\{
	\begin{aligned}
		& \chi_q(1 + \gamma_q^{(2)} \alpha^2), \ \text{if $q$ odd}
		\\& \chi_q \big(\alpha + \gamma_q^{(3)}\alpha^3\big), \ \text{if $q$ even}
	\end{aligned}
	\right.,
\end{equation}
where $\gamma_q^{(i)}$ represents the next higher-order terms in the corresponding expansions. From our discussion in Sec.~\ref{Sec:SM:g2}, we know that for both even and odd harmonics orders, where $\gamma_q^{(i)} = 0$, one has $g_q^{(2)}(0) = 1$ and  $g_q^{(2)}(0) = 3$, respectively, under low depletion conditions. Thus, a way to check whether the Gaussian description holds, both practically relevant given current experimental feasibility and illustrative, is to examine how the $g^{(2)}(0)$ function changes when these next higher-order terms are taken into account.

Broadly speaking, for $\langle \hat{a}^{\dagger n}\hat{a}^n \rangle$ we can write
\begin{equation}
	\langle \hat{a}^{\dagger n}\hat{a}^n \rangle
	= \int_{\mathbbm{R}} \dd\alpha
	\int_{\mathbbm{R}} \dd\beta\ 
	c(\alpha)c(\beta)
	\chi(\alpha)^n
	[\chi^*(\beta)]^n
	\braket{\beta}{\alpha}
	\prod_{q > 2} \braket{\chi_q(\beta)}{\chi_q(\alpha)},
\end{equation}
where for simplicity, we work under the assumption that the fundamental modes remain unchanged throughout the interaction. Given that the harmonic conversion efficiency in HHG is generally small,the dominant contribution to the integral above comes from the sharply peaked overlap $\braket{\beta}{\alpha}$, which for $\alpha,\beta \in \mathbbm{R}$ reduces to $\braket{\beta}{\alpha} = \exp[-(\beta-\alpha)^2/2]$, i.e., a Gaussian of unital variance centered at $\alpha = \beta$.~A natural way to evaluate the integral in $\beta$ is therefore via the stationary phase-approximation, which imposes $\alpha = \beta$ condition, yielding
\begin{equation}
	\langle \hat{a}^{\dagger n}\hat{a}^n \rangle
	\simeq \int_{\mathbbm{R}} \dd\alpha\
	c(\alpha)^2
	\abs{\chi(\alpha)}^{2n}
\end{equation}

Using this expression to compute the zero-time second-order autocorrelation function, we find
\begin{align}
	&g^{(2)}_{\text{odd}}(0)
	= \dfrac{1 + \gamma_{q,x}^{(2)}\sigma + \tfrac{9}{8} [\gamma_{q,x}^{(2)}\sigma]^2
		+ \tfrac{3}{8}[\gamma_{q,y}^{(2)}\sigma]^2+\tfrac{15}{16} ([\gamma_{q,x}^{(2)}\sigma]^3 +[\gamma_{q,y}^{(2)}\sigma]^3) + \tfrac{105}{256} ([\gamma_{q,x}^{(2)}\sigma]^4 +[\gamma_{q,y}^{(2)}\sigma]^4 ) + \tfrac{105}{128} [\gamma_{q,x}^{(2)}\sigma]^2 [\gamma_{q,x}^{(2)}\sigma]^2}{\big\{1 + \tfrac12 \gamma_{q,x}^{(2)}\sigma + \tfrac{3}{16} ([\gamma_{q,x}^{(2)}\sigma]^2 + [\gamma_{q,y}^{(2)}\sigma])\big\}^2}
	\\&
	g^{(2)}_{\text{even}}(0)
	= \dfrac{3+15\gamma_{q,x}^{(3)}\sigma + \tfrac{315}{8} [\gamma_{q,x}^{(3)}\sigma]^2
		+ \tfrac{945}{16}([\gamma_{q,x}^{(3)}\sigma]^3+ [\gamma_{q,x}^{(3)}\sigma][\gamma_{q,y}^{(3)}\sigma]^2
		+ \tfrac{10395}{256}([\gamma_{q,x}^{(3)}\sigma]^4+[\gamma_{q,y}^{(3)}\sigma]^4)
		+ \tfrac{10395}{128}[\gamma_{q,x}^{(3)}\sigma]^2[\gamma_{q,y}^{(3)}\sigma]^2 }{\big\{1 + \tfrac{3}{2}\gamma_{q,x}^{(3)}\sigma + \tfrac{15}{16}([\gamma_{q,x}^{(3)}\sigma]^2+[\gamma_{q,y}^{(3)}\sigma]^2)\big\}^2},
\end{align}
where we have written $\gamma_q^{(i)} = \gamma^{(i)}_{q,x} + i\gamma^{(i)}_{q,y}$. Taylor expanding both expressions up to first order in $\gamma^{(i)}_{j,q}\sigma$, we respectively arrive at
\begin{align}
	&g^{(2)}_{\text{odd}}(0)
	= 1 + \tfrac{1}{2}[\gamma^{(2)}_{q,x}\sigma]^2
	+ \tfrac{3}{4} [\gamma^{(2)}_{q,y}\sigma]^2 [\gamma^{(2)}_{q,x}\sigma] 
	- \tfrac{3}{16}[\gamma^{(2)}_{q,x}\sigma]^2[\gamma^{(2)}_{q,y}\sigma]^2 
	+ \mathcal{O}\big([\gamma^{(2)}_{q,x}\sigma]^3\big)
	+ \mathcal{O}\big([\gamma^{(2)}_{q,y}\sigma]^3\big),
	\\
	&g^{(2)}_{\text{even}}(0)
	= 3 + 6 \gamma^{(3)}_{q,x}\sigma 
	+ \mathcal{O}\big([\gamma^{(3)}_{q,x}\sigma]^2\big)
	+ \mathcal{O}\big([\gamma^{(3)}_{q,y}\sigma]^2\big).
\end{align}
These expressions set a requirement on the available squeezing strengths for the Gaussian description to remain accurate, corresponding to $\gamma_{q,x}^{(i)}\sigma \ll 1$, which, together with the correspondence $\bar{r}(r)$ established earlier, translates directly into an explicit bound on the physical squeezing parameter $r$ fo which the Gaussian description is expected to hold. In our case, we find $\gamma^{(i)}_{q,x}\sim 10^{-12}$ which, by imposing $\gamma_{q,x}^{(i)}\sigma \sim 1$, sets a limit on $r \simeq 14$, corresponding to $I_{\text{squ}} \sim 3.6 \times 10^{-5}$ a.u., consistent with the results of Fig.~\ref{Fig:Wigners}.

\section{Effect of Gaussian losses in the quantum optical state properties}\label{Sec:SM:Noise}
Throughout this work, macroscopic effects have been introduced phenomenologically through the $N$-scaling of the harmonic spectral response, under the assumption that the atomic contributions add coherently, i.e., that the relevant harmonics are phase-matched, reproducing the $N^2$ scaling of the coherent contribution to the harmonic yield~\cite{sundaram_high-order_1990,stammer_theory_2025}. However, phase-matching and propagation effects are expected to modify the quantum optical state and photon statistics of the generated harmonics~\cite{rivera-dean_propagation_2026,wang_generation_2026}.~In fact, Ref.~\cite{wang_generation_2026}, using a similar configuration to the one studied here but with the roles of the $\omega$ and $2\omega$ fields interchanged, found both theoretically and experimentally that the perturbative BSV field introduces a modification of the electronic wavepacket phase that affects the phase-matching conditions of the harmonics. Furthermore, they found that the optimal phase-matching can be controlled by adjusting the gas-jet pressure, with different sets of harmonics (even, odd and half-integer harmonics, i.e., those arising from the symmetries of the driving field) exhibiting different optimal pressure conditions, each within a relatively narrow range.~If such behavior also occurs in the configuration considered here, it suggests that the propagation conditions could in principle be tailored to optimize the phase matching of specific harmonic sectors.

However, even in the presence of perfect phase-matching, absorption in the medium can modify the correlations predicted in Sec.~\ref{Sec:Ent}.~Under negligible plasma generation, absorption can be represented by a lossy beam-splitter transformation that couples the harmonic field to an inaccessible environmental mode.~We denote by $\eta$ the fraction of the harmonic radiation lost to this ancillary mode, such that the amplitude transmission and loss coefficients are, respectively, $\sqrt{1-\eta}$ and $\sqrt{\eta}$. Assuming that the fundamental mode is essentially unaffected by absorption, i.e., that its interaction with the medium is accounted for only through the strong-field interaction, while all harmonic modes experience the same absorption, we can write the correlated state between the fundamental mode and the harmonic modes as
\begin{equation}
	\relaxket{\Tilde{\Phi}_{\text{even}}(t)}
	= \sqrt{\dfrac{(1+\bar{\chi}^2\sigma)^{\frac12}}{\pi \sigma}}
		\int_{\mathbbm{R}}\dd \alpha\
			\tilde{c}(\alpha)
			\relaxket{\bar{\chi}_{\mathsf{A}} \alpha,\bar{\chi}_{\mathsf{B}}\alpha\sqrt{1-\eta},\bar{\chi}_{\mathsf{B}}\alpha\sqrt{\eta}},
\end{equation}
where here $\mathsf{A}=\{2\}$, and the third mode represents the ancillary environmental mode into which part of the harmonic radiation is lost and which is subsequently traced out.

\begin{figure}[h!]
	\centering
	\includegraphics[width=1\textwidth]{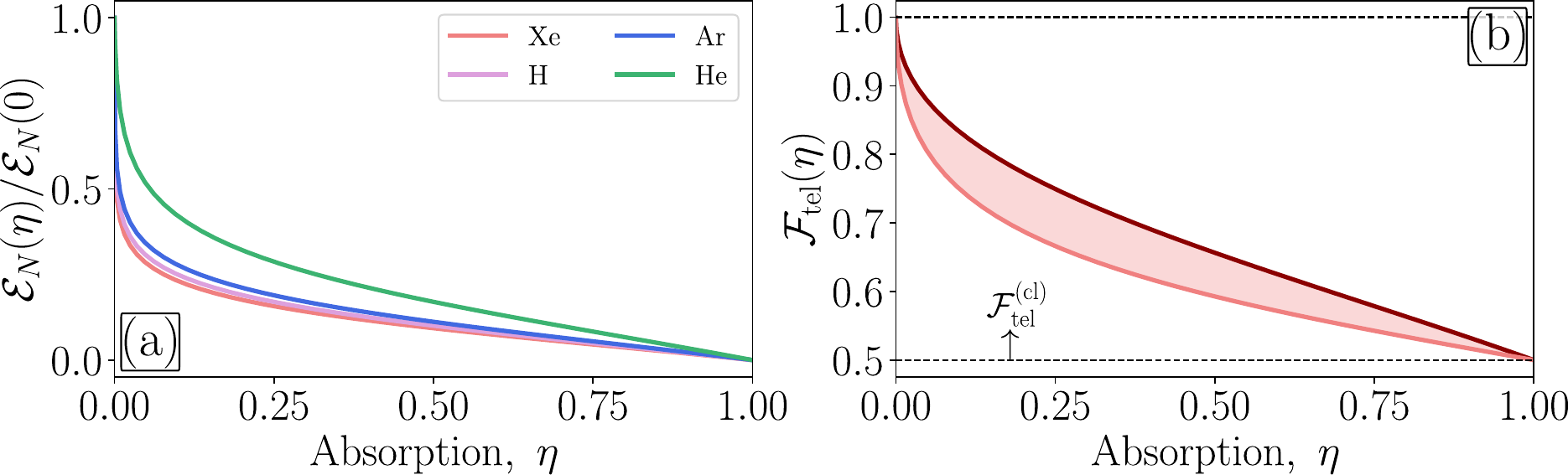}
	\caption{Effects of Gaussian absorption in the medium on (a) the logarithmic negativity, shown as the ratio between its value in the presence of absorption, $\mathcal{E}_N(\eta)$, and its value in the absence of absorption, $\mathcal{E}_N(0)$, and (b) the teleportation fidelity. For the latter, we focus on Xe, with the solid lines showing the upper and lower bounds of $\mathcal{F}_{\mathrm{tel}}(\eta)$; the other atomic species exhibit a qualitatively similar behavior. The same parameters as those in Figs.~\ref{Fig:Ent:DSI} and \ref{Fig:fid:teleport} have been used here.}
	\label{Fig:SM:Noise}
\end{figure}

Importantly, this loss process corresponds to a Gaussian channel and therefore preserves the Gaussian character of an initially Gaussian state. This makes the model particularly useful in the present context, as the Gaussian-state tools developed throughout this work remain directly applicable. Specifically, Eq.~\eqref{Eq:tot:Gamma} corresponds to the covariance matrix of the full state, with $\Gamma^{(\mathsf{B})}$ obtained from Eq.~\eqref{Eq:SM:Cov:matrix} upon the substitution $\bar{\chi}\rightarrow\bar{\chi}_{\mathsf{B}}\sqrt{1-\eta}$.
This shows that the contribution of the harmonic field to the covariance matrix scales with the transmitted intensity fraction $1-\eta$. Moreover, according to the calculations in Sec.~\ref{Sec:SM:g2}, the normalized second-order (auto)correlation function remains unchanged under this type of loss.

Figure~\ref{Fig:SM:Noise} displays the effects of Gaussian absorption on the entanglement properties of the state [panel (a)] and on the fidelity of teleportation [panel (b)].~Specifically, panel (a) shows the ratio between the logarithmic negativity obtained for a given value of $\eta$, $\mathcal{E}_N(\eta)$, and its value in the absence of absorption, $\mathcal{E}_N(0)$.~As observed, absorption significantly degrades the entanglement, with a particularly pronounced decrease at small $\eta$ that becomes progressively weaker as $\eta$ increases.~In the limiting case $\eta\rightarrow1$, corresponding to complete absorption of the harmonic field, the quantum optical state becomes fully separable.~Panel (b) shows the corresponding effect on the teleportation fidelity.~For this analysis, we focus on the case of Xe, with the solid lines displaying the upper and lower bounds of $\mathcal{F}_{\mathrm{tel}}(\eta)$; the other atomic species exhibit a qualitatively similar behavior.

\end{document}